\documentclass[fleqn,usenatbib]{mnras}

\usepackage{newtxtext,newtxmath}
\usepackage[normalem]{ulem}
\usepackage{graphics}
\usepackage{array}
\usepackage{hyperref}

\usepackage[T1]{fontenc}

\DeclareRobustCommand{\VAN}[3]{#2}
\let\VANthebibliography\thebibliography
\def\thebibliography{\DeclareRobustCommand{\VAN}[3]{##3}\VANthebibliography}

\usepackage{graphicx}	
\usepackage{amsmath}	

\title[]{Asteroseismological analysis of the polluted ZZ Ceti star G~29$-$38 with \emph{TESS}}

\author[Murat Uzundag et al.]{
Murat Uzundag$^{1}$\thanks{E-mail: muratuzundag.astro@gmail.com}, Francisco C. De Ger\'onimo$^{2,3,4,5}$, Alejandro H. C\'orsico$^{2,3}$, Roberto Silvotti$^{6}$, \newauthor 
Paul A. Bradley$^{7}$,  Michael H. Montgomery$^{8,9}$, M\'arcio Catelan$^{4,5}$, 
Odette Toloza$^{10,11}$, Keaton J. Bell$^{12}$, \newauthor 
S. O. Kepler$^{13}$,  Leandro G. Althaus$^{2,3}$,  Scot J. Kleinman$^{14}$, Mukremin Kilic$^{15}$, Susan E. Mullally$^{16}$, 
\newauthor Boris T. G\"ansicke$^{17}$, Karolina B\k{a}kowska$^{1}$,  Sam Barber$^{8,9}$,  Atsuko Nitta$^{18}$
\\
           $^{1}$Institute of Astronomy, Faculty of Physics, Astronomy and Informatics, Nicolaus Copernicus University in Toru\'n, Grudzi\k{a}dzka 5, PL-87-100 Toru\'n, Poland
           \\
           $^{2}$Grupo de Evoluci\'on Estelar y Pulsaciones. 
           Facultad de Ciencias Astron\'omicas y Geof\'{\i}sicas,
           Universidad Nacional de La Plata, Paseo del Bosque s/n, 1900, Argentina
           \\
           $^{3}$Instituto de Astrof\'isica de La Plata - CONICET
           \\
           $^{4}$Instituto de Astrof\'{\i}sica, Pontificia Universidad Cat\'olica de Chile, Av. Vicuña Mackenna 4860, 7820436 Macul, Santiago,  Chile
           \\
           $^{5}$Millennium Institute of Astrophysics, Nuncio Monse\~{n}or Sotero Sanz 100, Of. 104, Providencia, Santiago, Chile
           \\     
           $^{6}$INAF-Osservatorio Astrofisico di Torino, strada dell'Osservatorio 20, 10025 Pino Torinese, Italy
           \\
           $^{7}$XCP-6, MS F-699 Los Alamos National Laboratory, Los Alamos, NM 87545, USA
           \\
           $^{8}$Department of Astronomy, University of Texas at Austin, Austin, TX-78712, USA 
           \\
           $^{9}$McDonald Observatory, Fort Davis, TX-79734, USA
           \\
           $^{10}$Departamento de F\'isica, Universidad T\'ecnica Federico Santa Mar\'ia, Avenida Espa\~na 1680, Valpara\'iso, Chile
           \\
           $^{11}$Millennium Nucleus for Planet Formation, NPF, Valparaíso, Chile
           \\
           $^{12}$Department of Physics, Queens College, City University of New York, Flushing, NY-11367, USA
           \\
           $^{13}$Instituto de F\'isica, Universidade Federal do Rio Grande do Sul, 91501-970 Porto Alegre, RS, Brazil
           \\
           $^{14}$Astromanager LLC, Hilo, Hawaii, 96720, USA
           \\
           $^{15}$Homer L. Dodge Department of Physics and Astronomy, University of Oklahoma, 440 W. Brooks St., Norman, OK 73019, USA
           \\
           $^{16}$Space Telescope Science Institute, 3700 San Martin Drive, Baltimore, MD 21218, USA
           \\
           $^{17}$ Department of Physics, University of Warwick, Gibbet Hill, Coventry CV4 7AL, UK
           \\
           $^{18}$Gemini Observatory/NSF’s NOIRLab, 670 N. A’ohoku Place, Hilo, Hawai’i, 96720, USA
           \and
           }

\date{ Accepted 2023 September 7. Received 2023 September 7; in original form 2023 July 30
}

\pubyear{2023}

\begin{document}
\label{firstpage}
\pagerange{\pageref{firstpage}--\pageref{lastpage}}
\maketitle

\begin{abstract}

G\,29$-$38 (TIC~422526868) is one of the brightest ($V=13.1$) and closest ($d = 17.51$\,pc) pulsating white dwarfs with a hydrogen-rich atmosphere (DAV/ZZ Ceti class). It was observed by the {\sl TESS} spacecraft in sectors 42 and 56. The atmosphere of G~29$-$38 is polluted by heavy elements that are expected to sink 
out of visible layers on short timescales. The photometric {\sl TESS} data set 
spans $\sim 51$ days in total, and from this, we identified 56 significant pulsation
frequencies, that include rotational frequency multiplets.
In addition, we identified 30 combination frequencies in each sector.
The oscillation frequencies that 
we found are associated with $g$-mode pulsations, with periods spanning from $\sim$ 260 s to $\sim$ 1400 s. We identified 
rotational frequency triplets with a mean separation $\delta \nu_{\ell=1}$ of 4.67 $\mu$Hz and a quintuplet with a mean separation 
$\delta \nu_{\ell=2}$ of 6.67 $\mu$Hz, from which we estimated a rotation period of about $1.35 \pm 0.1$ days.
We determined a constant period spacing of 41.20~s for $\ell= 1$ modes and 22.58\,s for $\ell= 2$ modes. 
We performed period-to-period fit analyses and 
found an asteroseismological model with $M_{\star}/M_{\odot}=0.632 \pm 0.03$, $T_{\rm eff}=11\, 635\pm 178$ K, and  $\log{g}=8.048\pm0.005$ (with a hydrogen envelope mass of $M_{\rm H}\sim 5.6\times 10^{-5}M_{\star}$), in good agreement with the values derived from spectroscopy. 
We obtained an asteroseismic distance of 17.54 pc, which is in excellent agreement with that provided by {\sl Gaia} (17.51 pc).

\end{abstract}




\section{Introduction}
\label{intro}

DAV white dwarfs (WDs), also called ZZ Ceti stars, are pulsating hydrogen (H)-rich atmosphere  WDs with effective temperature in the range $10\,400 {\rm K}\lesssim T_{\rm eff}  \lesssim 13\,000$~K and surface gravities from $\log g\sim 7.5$ to $\sim 9$ \citep{2008ARA&A..46..157W, 2008PASP..120.1043F,2010A&ARv..18..471A,2019A&ARv..27....7C,2022PhR...988....1S, 2023MNRAS.522.2181K}. The discovery of pulsations in extremely low-mass white dwarfs extended these boundaries to cooler temperatures and lower surface gravities \citep{2013MNRAS.436.3573H}. ZZ Ceti stars constitute the most common class of pulsating WDs, with $\sim 500$ known members to date \citep{2016IBVS.6184....1B,2019A&ARv..27....7C,2020AJ....160..252V,2021ApJ...912..125G, 2022MNRAS.511.1574R}. These stars are multiperiodic pulsators, showing periods in the
range $100 \lesssim \Pi \lesssim 1400$\,s with amplitudes from 0.01 up to 0.3 magnitudes associated to spheroidal non-radial gravity ($g$) modes of low harmonic degree ($\ell= 1, 2$) and generally low to moderate radial order ($1 \lesssim k \lesssim 15$), excited by 
the convective-driving mechanism
\citep{1991MNRAS.251..673B, 1999ApJ...511..904G}. The existence of the 
red (cool) edge of the ZZ Ceti instability strip can be explained in terms of 
excited modes suffering enhanced radiative damping that exceeds 
convective driving, rendering them damped \citep{2018ApJ...863...82L}. 
In many cases, the ZZ Ceti pulsation spectrum exhibits rotational frequency splittings
\citep{1975MNRAS.170..405B}, 
which allows identifying modes and estimating the rotation period \citep[e.g.][]{2017ApJS..232...23H}.

While ground-based observations over the years have been extremely important in studying the nature of DAV stars \citep[e.g.,][]{1968ApJ...153..151L, 1990ApJ...361..309N, 2004ApJ...612.1052M,2008ARA&A..46..157W, 2008PASP..120.1043F,2021FrASS...8..229B}, observations from space have revolutionized the area of ZZ Ceti pulsations \citep[][]{2020FrASS...7...47C,2022BAAA...63...48C}. In particular, the {\sl K2} extension \citep{2014PASP..126..398H} of the {\sl Kepler} mission \citep{2010Sci...327..977B} allowed the discovery of outbursts in ZZ Cetis close to the red edge of the instability strip \citep{2015ApJ...809...14B,2018ApJ...863...82L}, and also the discovery that incoherent pulsations \citep{2017ApJS..232...23H} can give information about the depth of the outer convection zone \citep{2020ApJ...890...11M}. In addition, the Transiting  Exoplanet  Survey  Satellite  \citep[$TESS$;][]{2015JATIS...1a4003R} has allowed the discovery of 74 new ZZ Cetis \citep{2022MNRAS.511.1574R}. 

G\,29$-$38, also known as ZZ Psc, WD\,2326+049, EG\,159, and LTT\,16907, is a large-amplitude DAV star discovered to pulsate in 1974 by \cite{shulovkopatskaya1974}. Its variability was confirmed a year later by \citet{mcgrawrobinson1975}, showing from the beginning of its observation 
a complex and extremely variable pulsational spectrum. G\,29$-$38 has been the focus of numerous spectroscopic analyses. A compilation of $T_{\rm eff}$ and $\log g$ determinations can be found in Table~\ref{basic-parameters-targets}, based on the Montreal White Dwarf Database\footnote{\url{https://www.montrealwhitedwarfdatabase.org/}} \citep{2017ASPC..509....3D}. It is worth noting that the latest spectroscopic determinations of $T_{\rm eff}$ and $\log g$ are more reliable given that they account for corrections based on the three-dimensional hydrodynamical atmospheric simulations by \citet{2013A&A...559A.104T}.
The most recent spectroscopic determination is that of \cite{2020MNRAS.499.1890M} which gives $T_{\rm eff}= 11\,296\pm198$ K and $\log g= 8.02\pm0.03$.  This effective temperature 
places this star near the middle of the ZZ Ceti instability strip.
This star has been extensively studied for various combined properties that make it unique. G\,29$-$38 was the first single WD discovered to have an infrared excess \citep{1987Natur.330..138Z},  initially interpreted as arising from a brown dwarf companion. \cite{2003ApJ...584L..91J} showed that infrared excess can be due to an opaque flat ring of dust within the Roche region of the WD where an asteroid could have been tidally destroyed, producing a system reminiscent of Saturn’s rings. \citet{2018ApJ...866..108X} showed the flux of the infrared 10\,$\mu$m silicate feature increased by 10\% in less than 3 years, which they interpret to be caused by an increase in the mass of dust grains in the optically thin outer layers of the disk.
\cite{2020MNRAS.494.4591C} measured the polarization of optical light from G\,29$-$38 and searched for signs of stellar pulsation in the polarization data. Their data was limited and they were unable to demonstrate the impact of stellar oscillation.
The importance of fingering convection due to the accretion of surrounding material by
G\,29$-$38 was studied by \cite{2017A&A...601A..13W}. Recently, 
\cite{2022Natur.602..219C} detected X rays from 
G~29$-$38 based on  {\it Chandra}  observations and derived an accretion rate higher than estimates from past studies of the photospheric abundances. Finally, \cite{2023ApJ...944L..46E} revisited  
{\sl XMM Newton} data and also found X-ray emission at the location of G\,29$-$38,  with spectral properties of the source similar to those detected with Chandra observations.

Beyond these very interesting features related to the environment of the star, the main characteristic of G\,29$-$38 that is the focus of this paper is its pulsating nature and the possibility of probing its internal structure through asteroseismology. \citet{1997ASSL..214..445B} and \citet{Kleinman1998} explored the pulsation spectrum of G\,29$-$38 in great detail using a time-series photometry data set spanning 10 years, deciphering for the first time the complex and ever-changing pulsational spectra of a high-amplitude DAV star. 
G\,29$-$38 is reminiscent of cool DAVs located near the red edge of the ZZ Ceti instability strip. 
However, all the spectroscopic studies place the star closer to the middle of the instability strip.
\citet{Kleinman1998} detected 19 independent frequencies (not counting the non-central components of the rotational multiplets) with periods spanning the interval $110-1240$\,s, along with many combination frequencies. These authors plausibly suggested 
the harmonic degree and the radial order of 17 independent periods 
as being $\ell= 1$ and $k=1, 2, \cdots, 17$, and derived a mean constant period spacing of $\Delta \Pi(\ell= 1)\sim 47$ s. 
Further analyses of G\,29$-$38 were focused on time-resolved spectrophotometry. On the one hand, \cite{2000MNRAS.314..209V} identified six real modes and five combination frequencies. They measured small line-of-sight
velocities and detected periodic variations at the frequencies of five of 
the six real modes, with amplitudes of up to 5\,km/s \citep[in agreement with the expectations;][]{1982ApJ...259..219R},  conceivably due to
the $g$-mode pulsations. However, no velocity signals were
detected at any of the combination frequencies, thus confirming for the first 
time that the flux variations at the combination frequencies do not reflect global pulsations, but rather are the result of non-linear processes  
in the outer layers of the star. On the other hand, 
\cite{2000MNRAS.314..220C} derived the harmonic degree for the six modes detected by 
\cite{2000MNRAS.314..209V}, five of them (283\,s, 430\,s, 614\,s, 
653\,s and 818\,s) resulting from being dipole ($\ell= 1$) modes, and the mode with period 776\,s being a quadrupole ($\ell= 2$) mode. The presence and nature of the abundant linear combinations of frequencies in the pulsation spectrum of G\,29$-$38 were investigated in detail in a series of three articles by \citet{2000MNRAS.313..170V,2000MNRAS.313..179V} and \citet{2000MNRAS.313..185V}.
Subsequently, \cite{2003ApJ...589..921T} confirmed the measurements of the pulsation velocities detected by \cite{2000MNRAS.314..209V} and reaffirmed the fact that the frequency combinations and harmonics most likely result from non-linear mixing at 
the surface of the star and are not real modes that probe the interior, 
although they detected one combination mode with a significant velocity
signal. Later, \cite{thompson2008} presented optical time-series spectroscopy of G\,29$-$38 taken at the Very Large Telescope (VLT). These authors estimated
$\ell$ for 11 periods detected in this star, four of them being $\ell \neq 1$ modes. 
In particular, they derived an $\ell= 3$ or $\ell= 4$ value for the mode with 
period $\sim 353$\,s. 

The identification of the harmonic degree of a considerable number of modes of G\,29$-$38 prompted further model grid-based asteroseismological studies based on fits to individual periods. 
Specifically, three independent asteroseismological analyses of G\,29$-$38 were
carried out. The first one was that of \cite{2009MNRAS.396.1709C},  
based on the mean periods of the modes from different observations from 1985 to 1993\footnote{The list of periods employed by \cite{2009MNRAS.396.1709C} is not the same as that published by \cite{Kleinman1998} in their Table~3.},
assuming they are all $\ell= 1$ modes. They found a best-fit model with
$T_{\rm eff}= 11\,700$\,K, $M_{\star}= 0.665 M_{\sun}$, $M_{\rm He}=
10^{-2} M_{\star}$ and $M_{\rm H}= 10^{-8} M_{\star}$. The second 
asteroseismological analysis of this star was carried out by \cite{2012MNRAS.420.1462R}, 
based on the same list of periods as \cite{2009MNRAS.396.1709C}, 
but allowing $\ell$ to be 1 or 2. They found an asteroseismological model 
characterized by $T_{\rm eff}= 11\,471$\,K, 
$M_{\star}= 0.593 M_{\sun}$, $M_{\rm He}= 2.39 \times 10^{-2} M_{\star}$, and  
$M_{\rm H}= 4.67 \times 10^{-10} M_{\star}$. We note that, according to this 
asteroseismological model, 13 modes are $\ell = 2$ modes and only one is an $\ell= 1$ 
mode. The last asteroseismological analysis of this star was performed by \cite{2013RAA....13.1438C}, 
who employed the 11 periods and $\ell$ identifications
of \cite{thompson2008}. They found two equally valid asteroseismological models, 
one of them characterized by $T_{\rm eff}= 11\,900$\,K, $M_{\star}= 0.790\, M_{\sun}$, 
$M_{\rm He}= 10^{-2} M_{\star}$, and $M_{\rm H}= 10^{-4} M_{\star}$, 
and the other model with $T_{\rm eff}= 11\,250$\,K, $M_{\star}= 0.780\, M_{\sun}$, 
$M_{\rm He}= 3.16 \times 10^{-3} M_{\star}$, and $M_{\rm H}= 3.16 \times 10^{-6}M_{\star}$. 
These models are characterized by thick H envelopes, in contrast to the models 
of \cite{2009MNRAS.396.1709C} and \cite{2012MNRAS.420.1462R}, which have H envelopes 
several orders of magnitude thinner.

In this work,  we present new  {\sl TESS} observations of  G\,29$-$38.  We also 
perform a detailed asteroseismological analysis of this star on the basis of the 
fully evolutionary models of DA WDs computed by \cite{2010ApJ...717..897A} and \cite{2010ApJ...717..183R} and employed in our previous works on asteroseismology 
of ZZ Ceti stars \citep[][]{2012MNRAS.420.1462R,2013ApJ...779...58R, 2017A&A...599A..21D, 2017ApJ...851...60R, 2018A&A...613A..46D, 2019MNRAS.490.1803R, 2022MNRAS.511.1574R}. 
The paper is organized as follows. In Sect.~\ref{photometry} we describe the methods applied to obtain the pulsation periods of the target star. A brief summary of the stellar models of DA WD stars employed for the asteroseismological analysis of G\,29$-$38 
is provided in Sect.~\ref{evolutionary_models}. Section \ref{asteroseismological_modelling} is devoted to the asteroseismological modeling of the target star, including the search for a  possible uniform period spacing in the period spectrum, the derivation of the stellar mass using the period separation,  and the implementation of a  period-to-period fit with the goal of finding an asteroseismological model. Finally, in Sect.~\ref{conclusions}, we summarize our results.

\begin{table*}
  \caption{
  Effective temperature, surface gravity, spectral type, mass, luminosity, and cooling age measurements of G29-38 from different studies.
  }
  \begin{tabular}{cccccccc}
\hline
              &  $T_{\rm eff}$ & $\log g$ & Spectral  & Mass          & $\log (L_{\star}/L_{\sun})$      & Cooling age       & Reference \\
              &    [K]         & [cgs]    & type      & [$M_{\odot}$] &  & [Gyr]       &           \\
\hline

    & 11515$\,\pm\,$22 & 7.97$\,\pm\,$0.01  & DA  &  0.59  &   &     &   {\citet{2001A&A...378..556K}}  \\
    & 11600 & 8.05  & DAZ  &    &   &     &    {\citet{2003ApJ...596..477Z}}     \\
    & 11820$\,\pm\,$175 & 8.15$\,\pm\,$0.05  &   &  0.70$\,\pm\,$0.03  &  $-2.62$ & 0.55 & {\citet{2005ApJS..156...47L}} \\
    & 12100  & 7.90  & DAZ  &    &   &     &  {\citet{2005A&A...432.1025K}}       \\
    & 11600 & 8.10   & DAZ  &    &   &     &   {\citet{2006ApJ...646..474K}}      \\
    & 11485$\,\pm\,$80 & 8.07$\,\pm\,$0.02  &   &    &   &     &   {\citet{2009A&A...505..441K}}      \\
    & 12200$\,\pm\,$187 & 8.22$\,\pm\,$0.05  & DA  & 0.74$\,\pm\,$0.03    &   &     &   {\citet{2011ApJ...743..138G}}      \\
    & 12206$\,\pm\,$187 & 8.04$\,\pm\,$0.05  & DAZ  & 0.63$\,\pm\,$0.03   & $-2.50$  & 0.38     &   {\citet{2012ApJS..199...29G}}      \\
    & 11820$\,\pm\,$100 & 8.4$\,\pm\,$0.1  & DAZ  & 0.85   &   &     &    {\citet{2014ApJ...783...79X}}     \\
    & 12020$\,\pm\,$183 & 8.13$\,\pm\,$0.05  & DA  & 0.69$\,\pm\,$0.03   & $-2.58$  &     &   {\citet{2015ApJS..219...19L}}      \\
    & 11956$\,\pm\,$187 & 8.01$\,\pm\,$0.05  & DAV  & 0.61$\,\pm\,$0.03   &   &  0.38   &   {\citet{2016MNRAS.462.2295H}}      \\
    & 11240$\,\pm\,$360 & 8.00$\,\pm\,$0.03  & DAZV  & 0.60$\,\pm\,$0.03  & $-2.62$ $\,\pm\,$0.06  & 0.44$\,\pm\,$0.04   &   {\citet{2017AJ....154...32S}}      \\
    & 11315$\,\pm\,$180 & 8.02$\,\pm\,$0.06  & DA  &  0.62$\,\pm\,$0.08  &   &     &   {\citet{2017ApJ...848...11B}}      \\
    & 11295.9$\,\pm\,$198 & 8.02$\,\pm\,$0.03  & DAZ  &    &   &     &   {\citet{2020MNRAS.499.1890M}}      \\


\hline    
\label{basic-parameters-targets}
\end{tabular}
\end{table*}


\section{Photometric observations --- {\sl TESS}}
\label{photometry}

In this work, we investigate the pulsational properties of the well-known  DAV star G~29$-$38 using the high-precision photometry of {\sl TESS}  (see Table \ref{DAVlist}). G\,29$-$38 (TIC\,422526868), $G_{\rm mag}$= 13.06) was observed by 
{\sl TESS} in two sectors, including sector 42 (from 20 August to 16 September 2021) and sector 56 (from 01 September to 30 September 2022) in both 2 minutes and 20 seconds cadences. 
Using available magnitude values from the literature, we calculated the {\sl TESS}  magnitude of G\,29$-$38 as described by \citet{stassun2018} using the $\tt ticgen$\footnote{\url{https://github.com/TESSgi/ticgen}} tool, and found $T_{\rm mag}= 12.5$.  
The light curves were downloaded from The Mikulski Archive for Space Telescopes (MAST), which is hosted by the Space Telescope Science Institute (STScI)\footnote{\url{http://archive.stsci.edu/}} in FITS format. The light curves were processed by the Science Processing Operations Center (SPOC) pipeline \citep{Jenkins2016}.
We downloaded the target pixel files (TPFs) of G\,29$-$38 from the MAST archive with the Python package $\tt{lightkurve}$ \citep{lightkurve2020}. 
The TPFs feature an $11\times 11$ postage stamp of pixels from one of the four CCDs per camera that G\,29$-$38 was located on. 
To ascertain the degree of crowding and any other potential bright sources close to  G\,29$-$38, the TPFs were analyzed. Given that the {\sl TESS} pixel size is huge (21~arcsec), we checked any potential contamination through the $\tt CROWDSAP$ parameter, which provides the target flux to total flux ratio in the {\sl TESS} aperture. 
By examining the $\tt CROWDSAP$ parameter, which is provided in Table~\ref{DAVlist}, we were able to determine the level of contamination for G\,29$-$38. 
The $\tt CROWDSAP$ value is almost 1 for both sectors, suggesting that G\,29$-$38 is the source of the total flux measured by the {\sl TESS} aperture. 
The data have previously undergone processing with the \citet{Jenkins2016}  Pre-Search Data Conditioning Pipeline to eliminate common instrumental patterns. We initially extracted fluxes (``PDCSAP FLUX") and times in barycentric corrected Julian days ("BJD - 245700") from the FITS file. We then used a running 5$\sigma$ clipping mask to remove outliers.
We detrended the light curves to remove any additional low-frequency systematics that may be present in the data.
To do this, we applied a Savitzky–Golay filter with a three-day
window length computed with the Python package $\tt lightkurve$.
Finally, the fluxes were converted to fractional variations
from the mean, i.e. differential intensity $\Delta I/I$, and transformed to
amplitudes in parts-per-thousand (ppt). The ppt unit corresponds
to the milli-modulation amplitude (mma) unit\footnote{1 mma = 1/1.086 mmag = 0.1 \% = 1 ppt.} used in the past.
The final light curves of G\,29$-$38 from sector 42 (blue dots) and sector 56 (red dots) are shown in Fig.~\ref{Fig:LC_FT_42_56}.

\begin{table*}
  \caption{
  List of {\sl TESS} observations of G\,29$-$38,  including the {\sl TESS} input catalog number, {\sl TESS} magnitude, observed sectors,  date, $\tt CROWDSAP$,  and length of the runs. In addition, also listed are the temporal resolution, an average noise level of amplitude spectra, and detection threshold (which we define as the amplitude at $0.1\%$ of the FAP) which are obtained from the FT of the original and shuffled data (see text for details).
  }
  \begin{tabular}{cccccccccc}
\hline
TIC &  $T_{\rm mag}$ & Obs.   & Start Time        &  \tt{CROWDSAP} & Length & Resolution & Average Noise & $0.1\%$\,FAP \\
     &                & Sector & (BJD-2\,457\,000) &                & [d]    & $\mu$Hz    &  Level [ppt]  &   [ppt]  \\
\hline

422526868            & 12.5 & 42                   & 2447.6956 & 0.99  & 23.27  & 0.49 &  0.12     & 0.56     \\
                  &      & 56                   & 2825.2625       &   1.00  &    27.88     &  0.42    &  0.11  &  0.51     \\
\hline
\label{DAVlist}
\end{tabular}
\end{table*}

\subsection{Frequency analysis}

\begin{figure*}
        \includegraphics[width=\textwidth]{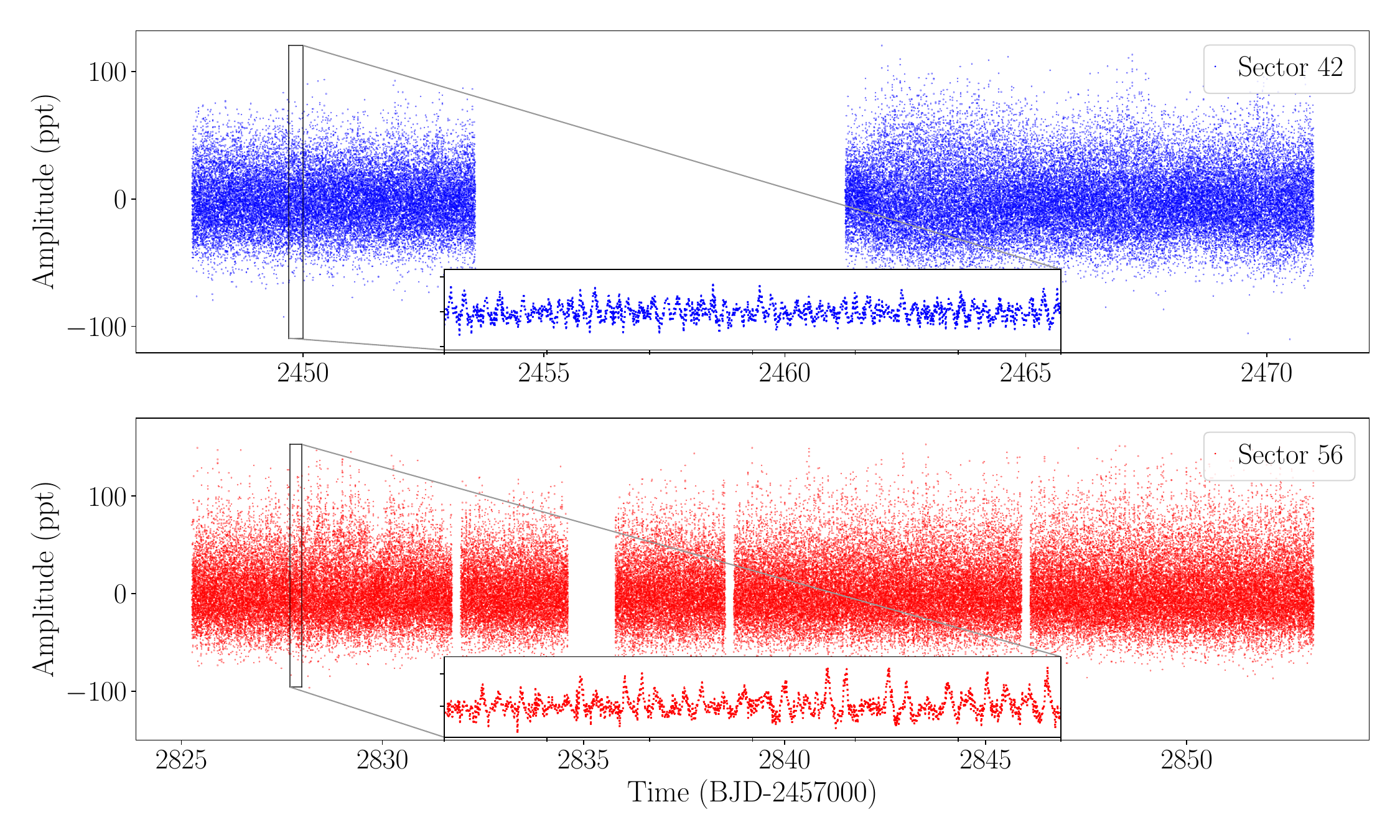}
	\includegraphics[width=\textwidth]{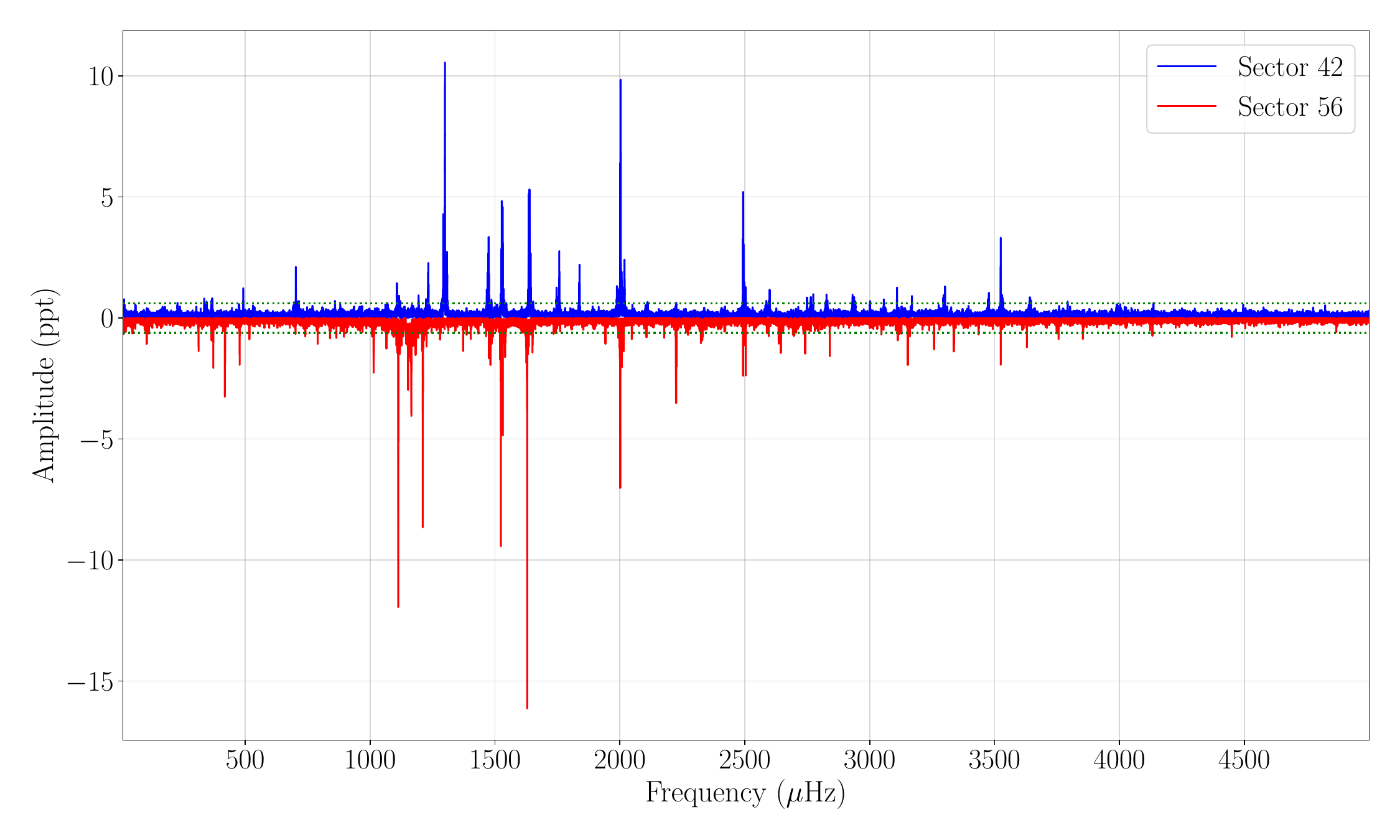}
    \caption{{Top:} Light curves of G\,29$-$38 from sector 42 (blue dots) and sector 56 (red dots). The insets are zoom of the light curves of 0.3 days to see the rapid variability.
    {Bottom:} Fourier transforms (FTs) of G\,29$-$38 computed from the sector 42 light curves (blue lines) and from the sector 56 light curves (red lines). The FT concentrates on the frequencies detected in the $g$-mode pulsation range. For the FT of sector 56, the amplitudes are inverted to improve clarity and comparison. The green dotted horizontal lines indicate the  $0.1 \%$ of the FAP significance threshold.}
    \label{Fig:LC_FT_42_56}
\end{figure*}

To carry out a thorough asteroseismic study, we aim at creating a comprehensive list of each of G\,29$-$38's 
independent frequency and linear combination frequencies observed. In order to examine the periodicities in the data and determine the frequency of each pulsation mode, together with its amplitude and phase, Fourier transforms (FT) of the light curves were obtained. 
In Fig.~\ref{Fig:LC_FT_42_56}, we depict the FT of sector 42 with blue lines and the FT of sector 56 with red lines.

We used our customized tool for a prewhitening procedure, which uses a nonlinear least square (NLLS) algorithm to fit each pulsation frequency in a waveform $A_i \sin(\omega_i\ t + \phi_i)$, with $\omega_i=2\pi/P_i$, and $P_i$ the period. In addition, we make use of two different publicly available tools of $\tt Period04$\footnote{\url{http://www.period04.net/}} \citep{LB2005} and $\tt Pyriod$\footnote{\url{https://github.com/keatonb/Pyriod}} to identify the frequency of each pulsation mode.
We fitted each frequency that appears above the 0.1\%\ false alarm probability (FAP).
The FAP level was calculated by reshuffling the light curves 1000 times as described in \citet{Kepler1993}. 
The temporal resolution of the data is about 0.49 $\mu$Hz ($1/T$, where $T$ is the data time length, which is 23.27 d) for sector 42, while the temporal resolution for sector 56 is around 0.42 $\mu$Hz as the star was observed during 27.88\,d. 
Table~\ref{DAVlist} lists all relevant information regarding the FT, including the average noise level and the $0.1\%$ FAP level of each dataset.
For all the peaks that are above the accepted threshold and up to the frequency resolution of the particular dataset, we performed a non-linear least squares (NLLS) fit. 
This iterative process has been done starting with the highest peak until there is no peak that appears above $0.1 \%$ of the FAP significance threshold. However, G\,29$-$38 exhibits significant amplitude, frequency and/or phase variations over the duration of each run, resulting in an excess of power in the FT after pre-whitening.
We carefully analyzed all frequencies that still had any excess power over the threshold after pre-whitening to see whether there was a close-by frequency within the frequency resolution, and only the highest amplitude frequency was fitted and pre-whitened in such cases.

\subsection{Frequency solution from sector 42}

The frequency spectrum from sector 42 shows a rich content of peaks between $\sim 220$ and $4150\,\mu$Hz. 
We employed NLLS fits to determine the values of around 60 signals above the detection limit of 0.1\%\,FAP = 0.56\,ppt. 
Considering that the median noise level over the whole FT is 0.12 ppt, the observed frequencies located between 200 and $4000\,\mu$Hz have S/N between 5 and 84.

All pre-whitened frequencies for G~29$-$38 including only sector 42 are given in Table~\ref{table:S42_Flist}, showing frequencies (periods) and amplitudes with their corresponding uncertainties and the S/N ratio. 

\begin{figure}
        \includegraphics[width=\columnwidth]{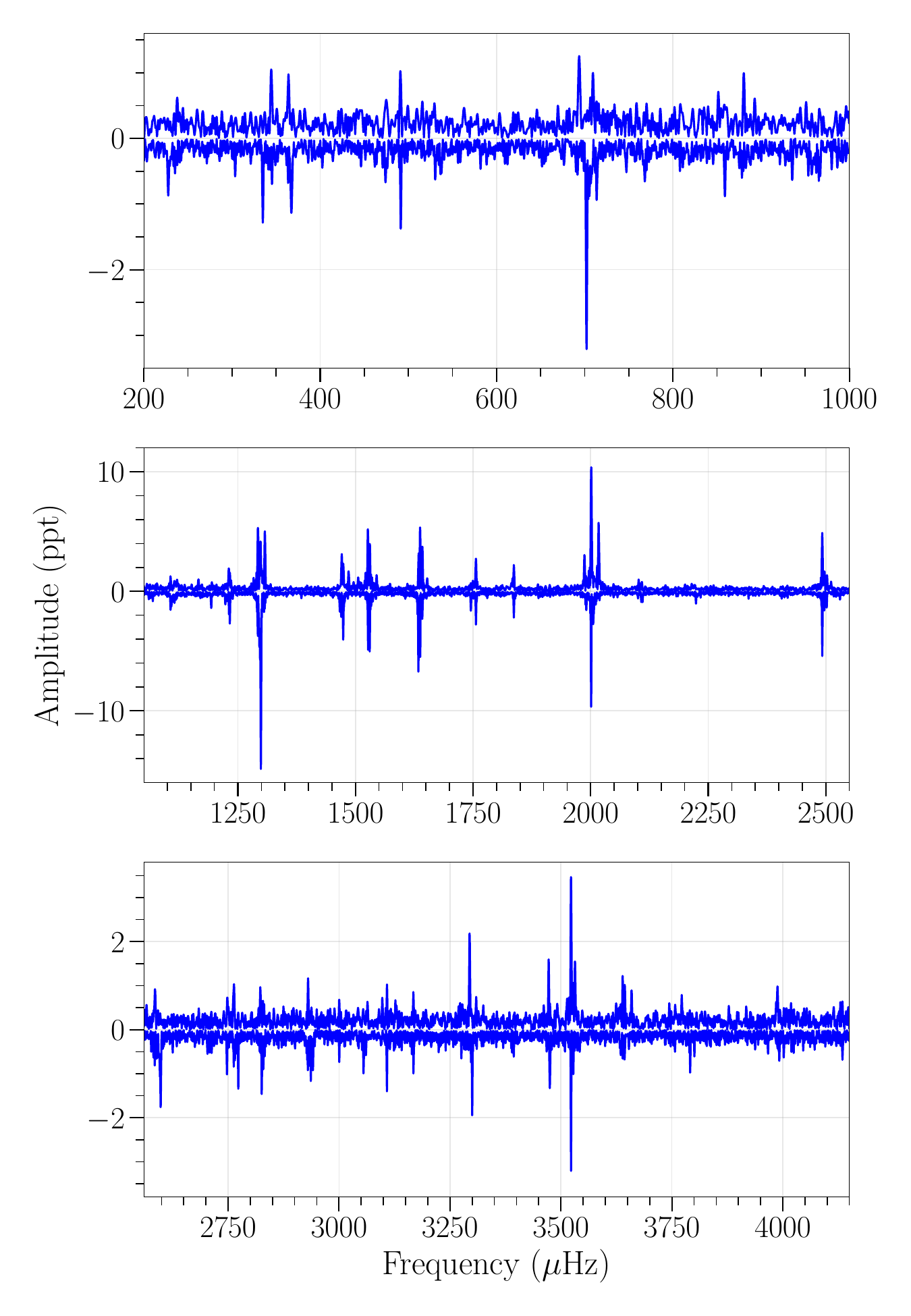}
    \caption{
    FT of the first and second half of sector 42, displaying amplitude changes during the 23\,d observation. The amplitudes of the second half are inverted for improved clarity and comparison.}
    \label{fig:FT_42_1st_2nd}
\end{figure}

In sector 42, there is a 7.67 days gap in the light curve as can be seen in Fig.~\ref{Fig:LC_FT_42_56}. 
We calculated the FT of each of the two halves of the light curve. The first chunk lasts for approximately 5.88 days, while the second chunk covers 9.72 days. 
Fig.~\ref{fig:FT_42_1st_2nd} shows the FT of the first half and the second half in three panels. In the FT of the second half of the light curve, the amplitudes are inverted for clarity. The  upper panel of Fig.~\ref{fig:FT_42_1st_2nd} displays  the short frequency region showing a notable difference in the peak located at 700\,$\mu$Hz. In the second half of sector 42, the amplitude increases by a factor of two at 700\,$\mu$Hz. The frequencies at 350 and 900\,$\mu$Hz show both amplitude and frequency changes.   
The second panel of Fig.~\ref{fig:FT_42_1st_2nd} displays the peaks at 1300 and 2000$\,\mu$Hz where a substantial difference was seen. Particularly, the peak at 2000\,$\mu$Hz displays a triplet pattern; however, it gradually disappears at the FT of sector 42's second half. Similarly, in the second half of sector 42, the amplitude increases by a factor of two at 1300\,$\mu$Hz, and the side components of the main peak disappear. We observed significant changes in the amplitudes in the long frequency range, which is depicted in the third panel of Fig.~\ref{fig:FT_42_1st_2nd}, notably beyond 3250\,$\mu$Hz, where all of the peaks exhibit amplitude variations.

\subsection{Frequency solution from sector 56}

The FT of the light curve from sector 56 reveals a plethora of peaks between 100 and 4450\,$\mu$Hz. 
In total, 66 frequencies were detected above the detection limit of 0.1\%FAP= 0.51\,ppt, and were extracted from the light curve through an NLLS fit. 
The median noise level is 0.11\,ppt and the detected frequencies have S/N values spanning from about 6 to 149.
Table~\ref{table:S56_Flist} contains all pre-whitened frequencies for G\,29$-$38, including only sector 56, and provides frequencies (periods) and amplitudes with their associated uncertainties and the S/N ratio. 

\subsection{Combination frequencies}

Combination frequencies are observed in the FTs of many $g$-mode pulsators, including low amplitude pulsating stars such as variable hot subdwarf B  and WD stars, and low to large amplitude pulsators such as $\gamma$~Dor stars and slowly pulsating B stars (SPBs). \citet{2022ARA&A..60...31K} reviewed the details and feasibility of combination frequencies across the Hertzsprung-Russell Diagram of pulsating stars. Combination frequencies have been detected in several classes of pulsating WD stars, including DOVs, DBVs, and DAVs. The precise numerical correlations between combination frequencies and their parent frequencies are used to identify them. The frequency combination peaks are not self-excited, but rather result from nonlinear processes linked with the surface convection zone and can be used to infer the latter's thermal response timescale \citep{2005ApJ...633.1142M}.

Both sectors have numerous combination frequencies.  {\sl TESS} observations resolve around 30 combination frequencies per sector. 
A complete list of combination frequencies is provided in Tables~\ref{table:S42_Flist} and \ref{table:S56_Flist}, for sectors 42 and 56, respectively.  
In order to count as a combination frequency, we made two assumptions.
First, we assumed that linear combinations have lower amplitudes than their parent frequencies. Second, we designated a combination frequency if the difference between the parent and combination frequency was within the frequency resolution of $\sim$ 0.5 $\mu$Hz.

\begin{figure}
        \includegraphics[width=\columnwidth]{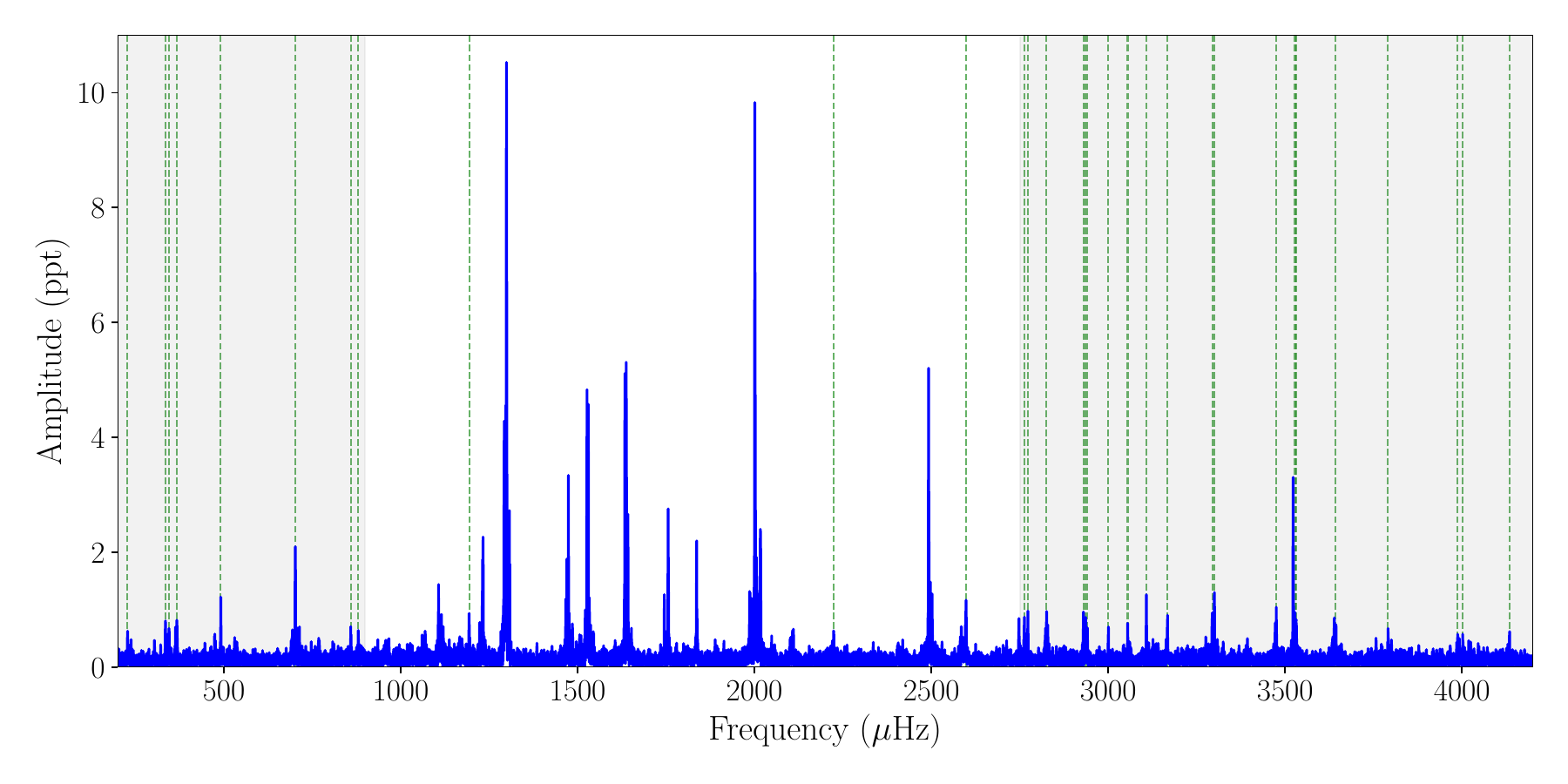}
    \caption{FT of sector 42, showing the location of combination frequencies (vertical dotted green lines). The low ($\leq$ 900\,$\mu$Hz) and high ($\geq$ 2750\,$\mu$Hz) frequency regions, which are depicted as grey-shaded areas, are where the majority of the combination peaks are found. 
    }
    \label{Fig:comb_42}
\end{figure}

In the case of sector 42, we detected 30 combination frequencies, and  $\sim$93\% of which were located either in the short- ($\leq$ 800\,$\mu$Hz) or long-frequency ($\geq$ 2750\,$\mu$Hz) regions, as illustrated with grey shaded regions in Fig.~\ref{Fig:comb_42}. In this plot, the location of each combination frequency is shown with a vertical dashed green line. We detected only two combination frequencies out of these regions, at 1193 and 2223\,$\mu$Hz.
While the mean S/N of the parent peaks corresponds to 24, the mean of S/N of the combination frequencies corresponds to 7.

\begin{figure}
        \includegraphics[width=\columnwidth]{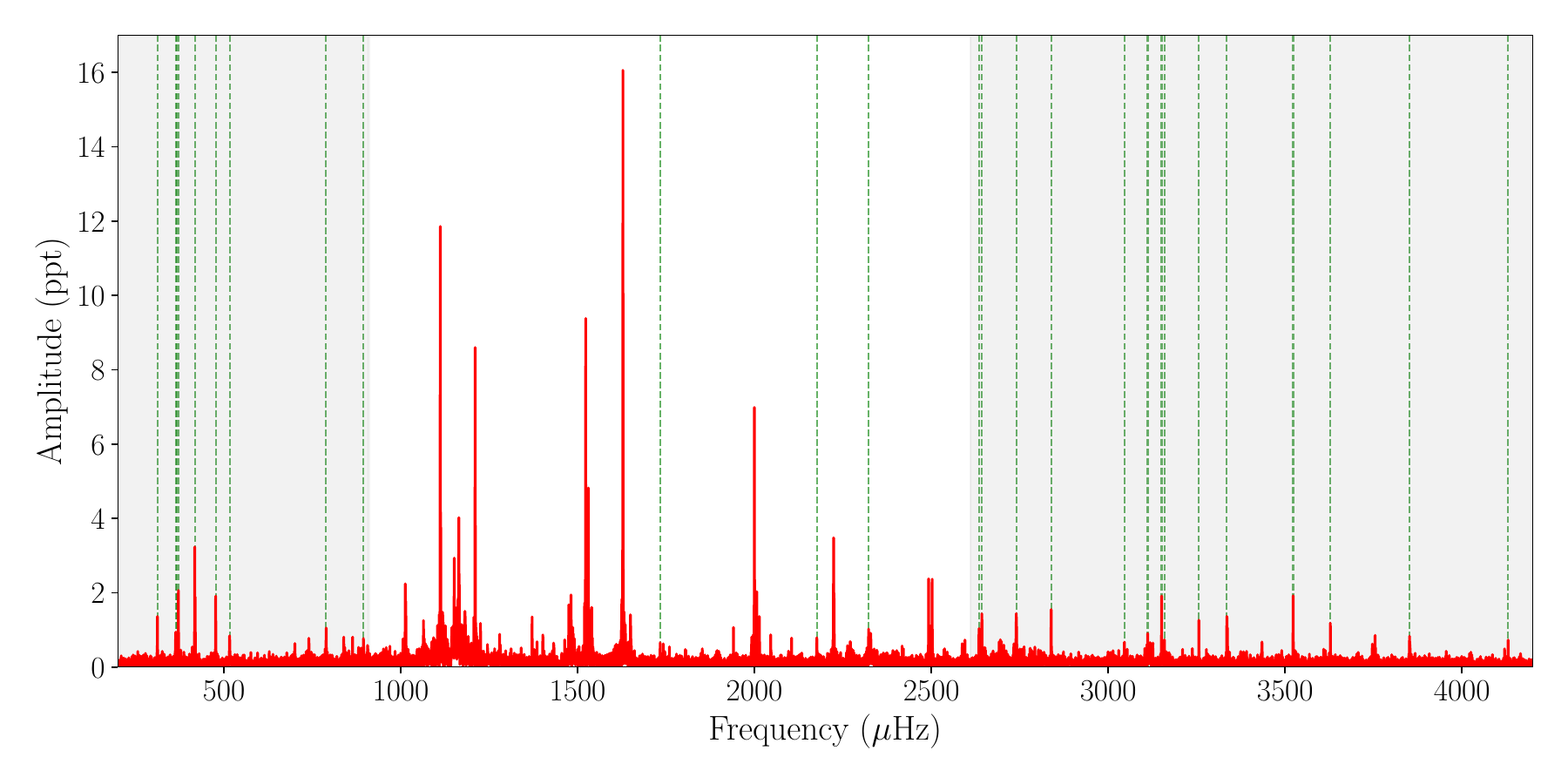}
    \caption{FT of sector 56, showing the location of combination frequencies (vertical dotted green lines). The low ($\leq$ 900\,$\mu$Hz) and high ($\geq$ 2610\,$\mu$Hz) frequency regions, which are depicted as grey-shaded areas, are where the majority of the combination peaks are found. 
    }
    \label{Fig:comb_56}
\end{figure}

As seen in the grey shaded regions of Fig. \ref{Fig:comb_56}, we identified 29 combination frequencies for sector 56, and around 90 percent of them were found in the short- ($\leq$ 900\,$\mu$Hz) and long-frequency ($\geq$ 2610\,$\mu$Hz) regions.
Out of these two areas, only three frequencies at 1733, 2176, and 2322\,$\mu$Hz were detected. The precise location of each combination frequency is presented in Fig.~\ref{Fig:comb_56} with the vertical dashed green line.
The mean S/N of combination frequencies in this case, however, equates to 11, whereas the mean S/N of parent peaks corresponds to 27.

\subsection{Mode identification}

To constrain the internal structure of G\,29$-$38 with asteroseismology, our primary goal is to identify the modes of the observed pulsations.
The nonradial pulsation modes are characterized by three quantized numbers, $k, \ell$, and $m$, where $k$ represents the number of radial nodes between the center and the surface, $\ell$ the number of nodal lines on the surface, and $m$ the azimuthal order, which denotes the number of nodal great circles connecting the star's pulsation poles. 
To identify the pulsational modes of G\,29$-$38, we applied two methods, namely rotational multiplets, and asymptotic period spacing, as discussed in the following sections.

\subsection{Rotational multiplets}

Rotational multiplets can be used to ascertain the rotation period and pinpoint the pulsation modes in rotating stars when nonradial oscillations are present \citep[][and references therein]{1980tsp..book.....C,1989nos..book.....U,2010aste.book.....A,2015pust.book.....C}. 

The eigenfrequencies of harmonic degree $\ell$ break into $2{\ell}+1$ components that differ in azimuthal ($m$) number owing to slow stellar rotation, which is a well-known feature of nonradial stellar pulsations. When the rotation is slow and rigid, the frequency splitting is $\delta \nu_{\ell, k, m}= m\ (1-C_{\ell,k})\ \Omega_{\rm R}$, $\Omega_{\rm R}$ being the rotational angular frequency of the pulsating star and $m= 0, \pm 1, \pm 2, \cdots, \pm \ell$ \citep[e.g.][]{1989nos..book.....U}.
The slow rotation requirement means that $\Omega_{\rm R} \ll \nu_{\ell, k}$. The $C_{\ell, k}$ constants are the Ledoux coefficients  \citep{1958HDP....51..353L}, which may be calculated as $C_{\ell,k} \sim [\ell(\ell+1)]^{-1}$ in the asymptotic limit of high-order $g$ modes ($k \gg \ell$). 
In the asymptotic limit, $C_{1,k}\sim  0.5$ and $C_{2,k} \sim 0.17$ in the case of $\ell= 1$ and $\ell= 2$ modes, respectively. Multiplets in the frequency spectrum of a pulsating WD are highly valuable for identifying the harmonic degree of the pulsation modes, in addition to enabling an estimate of the rotation period of nonradial pulsating stars.
Rotational multiplets have been found in all classes of pulsating WDs, including GW Vir, DBV, and DAV stars, with calculated rotation periods ranging from an hour to a few days.
The method's application and recent examples can be found in \citet{2017ApJS..232...23H,2022A&A...668A.161C,2022MNRAS.513.2285U,2022ApJ...936..187O} and \citet{2023MNRAS.518.1448R}.

\begin{figure*}
	\includegraphics[width=\textwidth]{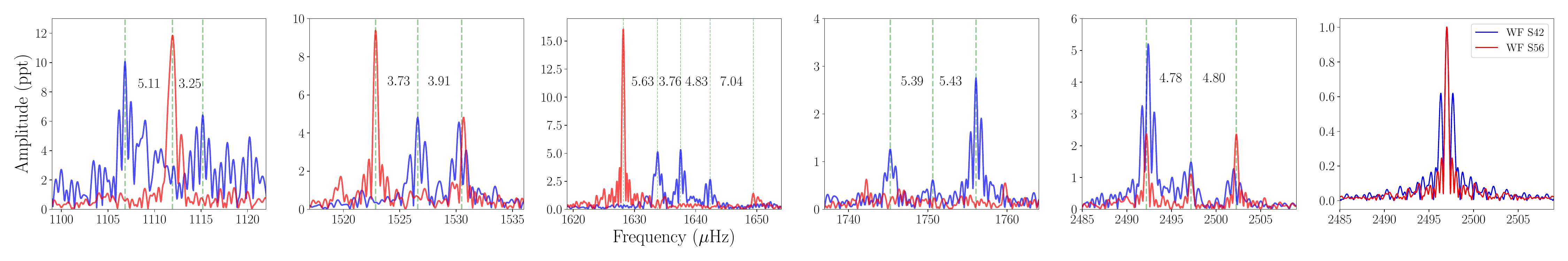}
    \caption{The amplitude spectra of G\,29$-$38 that is calculated based on sector 42 (blue lines) and sector 56 (red lines) showing rotational triplets. The first sub-figure presents the window 
function of G\,29$-$38 with the same color code centered at 1526.59\,$\mu$Hz for comparison.
    The remaining sub-plots display rotational splittings in four different regions in the amplitude spectra, with an average splitting of $\delta\nu = 4.57\,\mu$Hz. }
    \label{triplet}
\end{figure*}

Since the FTs from both sectors show different structures, we interpreted each FT separately to search for rotational triplets and quintuplets. 
In the FT of sector 42, we found three distinct triplets, whose central components ($m\,=\,0$) are located at 1637.552, 1750.641, and 2497.176\,$\mu$Hz, with an average splitting of $\delta\nu = 4.83\,\mu$Hz. 
We depict these three triplets in the third to fifth panel of Fig.~\ref{triplet}, along with the window function (sixth panel) and a doublet at 1526.59 and 1530.251\,$\mu$Hz (second panel). 
Among these three multiplets, the only triplet that is found in the FT of sector 56, is located at 2497.176\,$\mu$Hz. This triplet was also detected by \citet{Kleinman1998}. The other ones are either completely absent or incomplete, showing two components in the FT of sector 56. For instance, the triplet with central components at 1637.552 and 1750.641\,$\mu$Hz is absent. However, two additional peaks appear at 1628.166 and 1649.424\,$\mu$Hz, making the interpretation difficult.
These two peaks might be independent of the triplet that is resolved in the FT of sector 42, or they could be interpreted as rotational quintuplets. However, in that case, the splittings are inconsistently spanning from 3.76\,$\mu$Hz to 7.04\,$\mu$Hz. 
Thus, based on the FT from sector 42, we assessed 1633.792, 1637.552, and 1642.383\,$\mu$Hz as rotational triplets. The doublet detected at 1526.59 and 1530.251\,$\mu$Hz becomes complete when the FT from sector 56 is included.  This region was also resolved in the dataset provided by \citet{Kleinman1998}, indicating that rotational multiples may exist, although their data were equally inconclusive. Lastly, in the first panel of Fig.~\ref{triplet}, we showed another candidate at 1106.833, 1111.944, and 1115.196\,$\mu$Hz, with a splitting of 5.11 and 3.25\,$\mu$Hz, respectively. All these candidates are listed in Table~\ref{table:G29_42_ModeID} with their rotational splittings ($\delta\nu$).
Overall, the splitting for $\ell = 1$ modes from 3.73~$\mu$Hz to 5.43~$\mu$Hz provides a mean rotation period range between 1.07 and 1.55 days.
If we include all these five candidates as potential rotational multiples, then the average splitting is $\delta\nu = 4.57\,\mu$Hz. This provides a rotation period for G\,29$-$38 of $\sim$1.24 days, which aligns with what was reported by \citet{Kleinman1998}.

\begin{figure*}
	\includegraphics[width=\textwidth]{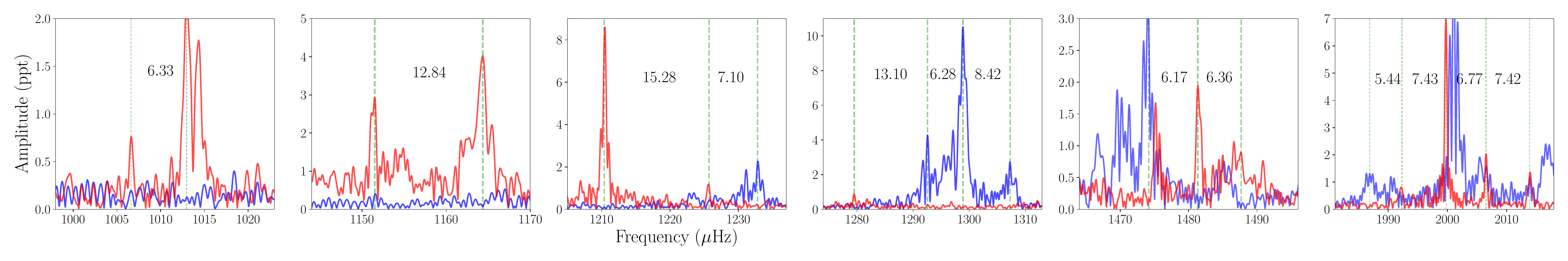}
    \caption{Same as in Fig.~\ref{triplet} to display rotational quintuplets. The sub-plots present rotational splittings in three different regions in the amplitude spectrum, with an average splitting of $\delta\nu = 6.66\,\mu$Hz for $\ell = 2$ modes. See text for more details. }
    \label{quintuplet}
\end{figure*}

Once we determine the $\ell =1$ triplets, we may look for modes with higher modal degrees. According to the previously mentioned Ledoux formula, the splitting in $\ell =2$ quintuplets is $\sim 1.67$ times larger than in $\ell =1$ triplets, which range from 3.73 to 5.43\,$\mu$Hz. In the case of  $\ell =2$  quintuplets, higher modal degree modes will have even larger splittings, ranging from  6.23 to 9.07\,$\mu$Hz.
The structure of the candidates of rotational quintuplets is complex, as shown in Fig.~\ref{quintuplet}, probably due to the detected amplitude modulation. We found only one candidate with a complete structure showing 5 azimuthal orders from 1986.868 ($m=-2$) to 2016.62~$\mu$Hz ($m=+2$) and average splitting of $\delta\nu = 6.77\,\mu$Hz, which is shown in the sixth panel in Fig.~\ref{quintuplet}. None of the remaining candidates show the complete structure, and the components vary sector by sector as in the case of dipole multiplets. The splittings for $\ell = 2$ modes (except for a quintuplet at 1999.742\, $\mu$Hz) span from 6.17 to 8.42\,$\mu$Hz with an average splitting of $\delta\nu = 6.81\,\mu$Hz. 
Taking into account all these six candidates as quintuplets, the average splitting is $\delta\nu = 6.66\,\mu$Hz. This provides a mean rotation period for G~\,29$-$38 of $\sim$1.45 days.  


\subsection{Asymptotic period spacing}
\label{APS}

The periods of $g$-modes with consecutive radial order are roughly evenly separated \citep[e.g.][]{1990ApJS...72..335T} in the asymptotic limit of high radial orders  ($k \gg \ell$), being the constant period spacing dependent on the harmonic degree:
\begin{equation}\label{eq:1}
\Delta \Pi_{\ell}^{\rm a} = \frac{{\Pi}_{0}}{\sqrt{\ell(\ell+1)}},
\end{equation} 
$\Pi_{0}$ being a constant value defined as
\begin{equation}\label{eq:2}
\Pi_{0}=  \frac{2 \pi^2}{\left[ \int_{r_1}^{r_2}
\frac{N}{r} dr \right]}, 
\end{equation} 
where $N$ is the Brunt-V\"ais\"al\"a frequency.  
The asymptotic period spacing given by Eq.\~(\ref{eq:1}) is very close to the computed period spacing of $g$-modes in chemically homogeneous stellar models without convective regions \citep{1980ApJS...43..469T}. In the case of DAVs, the asymptotic period spacing (and of course, also the average of the computed period spacings) is a function of the stellar mass, the effective temperature, and the thickness of the H envelope, with similar degrees of sensitivity to each parameter \citep{1990ApJS...72..335T}.
This implies that measuring a period spacing in G\,29$-$38 can be useful for identifying the harmonic degree of the observed frequencies, but caution should be exercised in using it to derive an estimate of stellar mass, due to the simultaneous dependence of the period spacing on $M_{\star}$, $T_{\rm eff}$, and $\log(M_{\rm H}/M_{\star})$. The latter does not happen in the case of DBVs and GW~Vir stars, since for them, 
the period spacing is basically dependent only on $T_{\rm eff}$ and $M_{\star}$ \citep{2021A&A...645A.117C,2022A&A...659A..30C, 2022A&A...668A.161C}. That said, 
however, in Sect.~\ref{aps} we will show that it is still feasible to derive a 
\emph{range} of stellar mass values for G\,29$-$38 on the basis of the observed period 
spacing, disregarding the exact value of $T_{\rm eff}$ and $M_{\rm H}$.

\begin{figure}
	\includegraphics[width=\columnwidth]{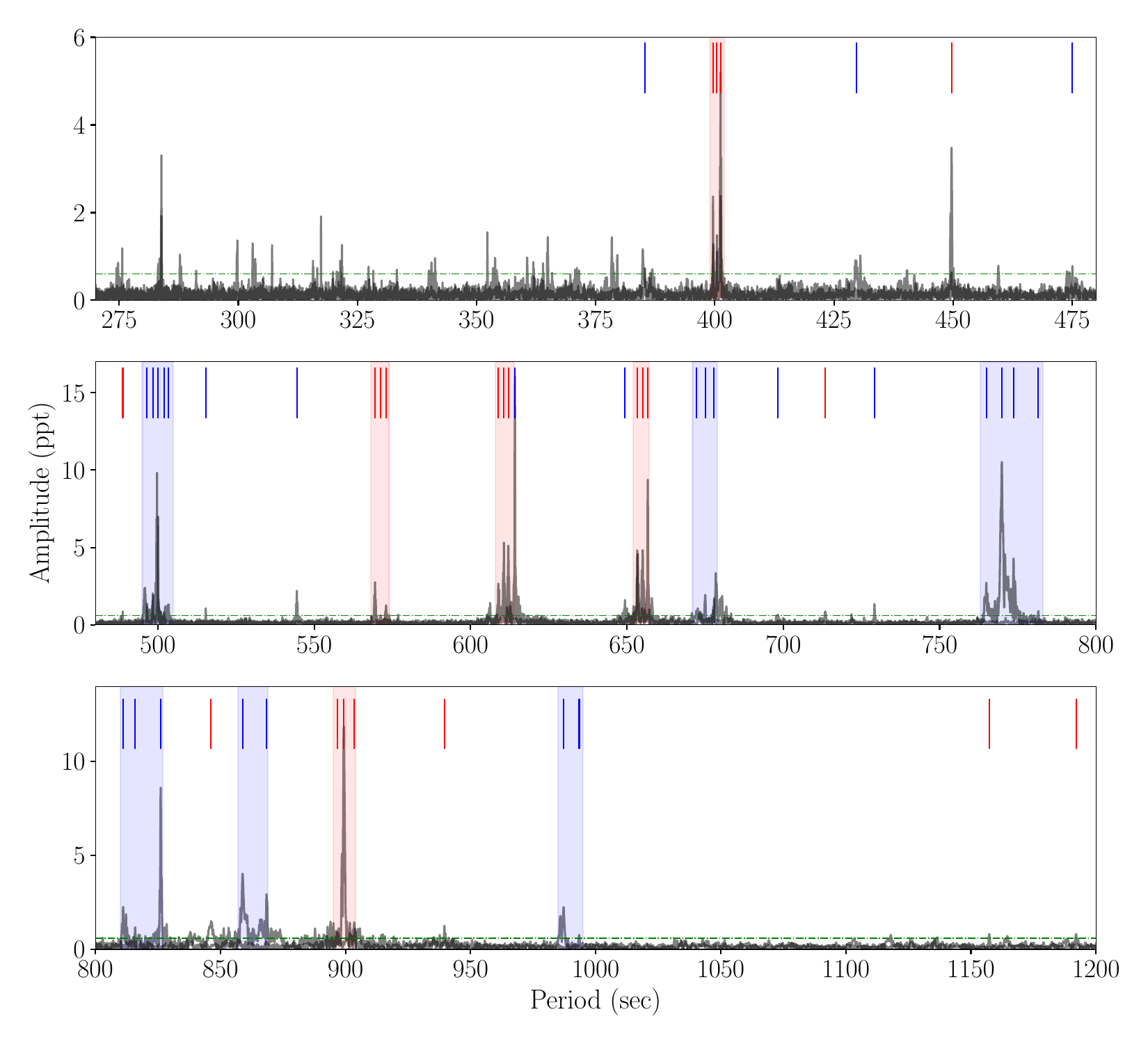}
    \caption{ Pulsation spectrum of G\,29$-$38 in terms of the periods based on combined sectors 42 and 56. 
The vertical red lines (blue lines) indicate the location of $\ell=1$ ($\ell= 2$) 
periods that make up the patterns of constant dipole and quadrupole period spacings.
The vertical red (blue) shaded regions correspond to the potential rotational multiples (quintuplets) that are zoomed in Figs. \ref{triplet} and \ref{quintuplet} in the frequency domain. 
The horizontal blue lines show the confidence level of 0.1\% FAP which is calculated based on sector 42. 
    Detected modes are labeled in Table~\ref{table:G29_42_ModeID}. }
    \label{FTG29_S42_ModeID}
\end{figure}

\begin{figure}
	\includegraphics[width=\columnwidth]{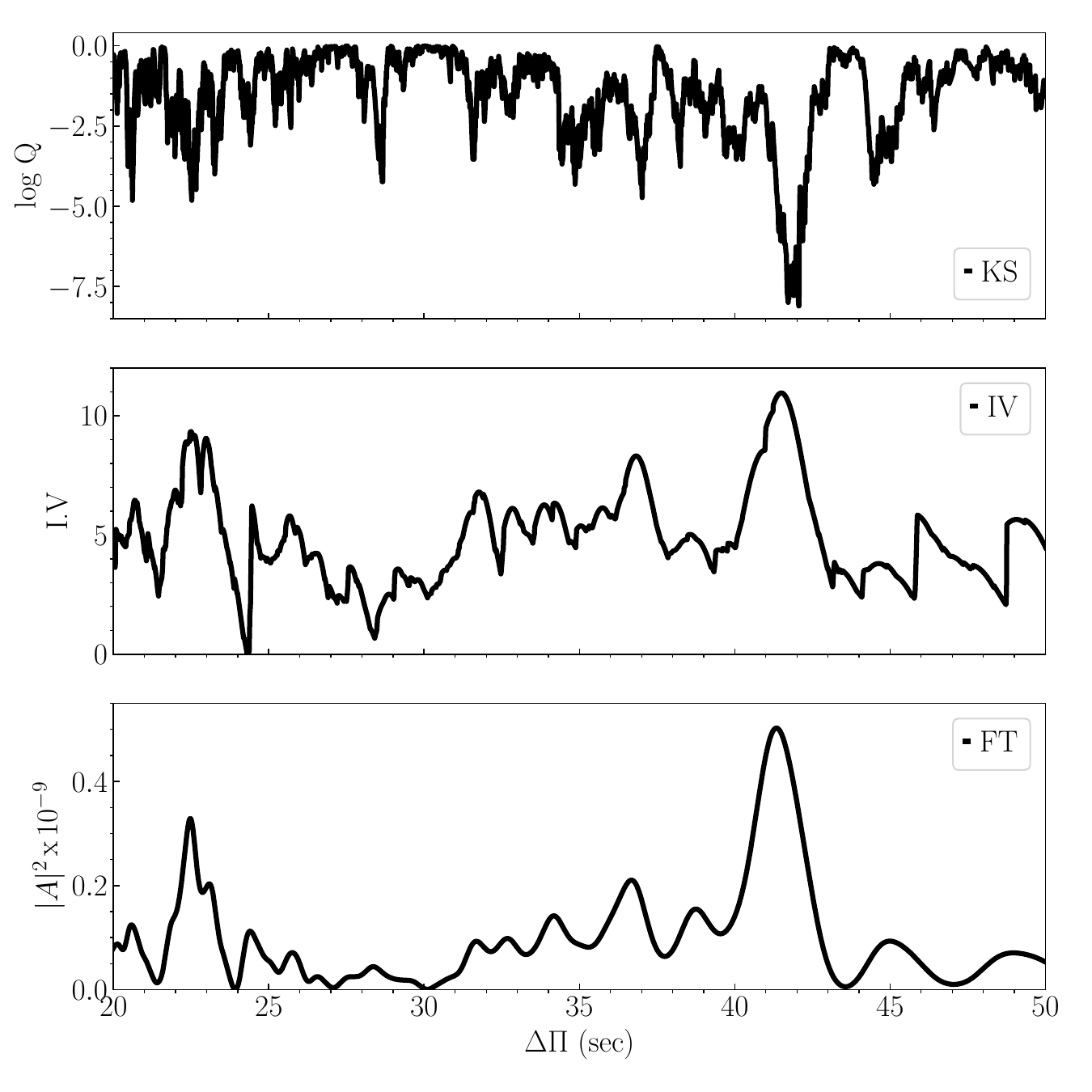}
    \caption{K-S (upper panel), I-V (middle panel), and F-T (bottom
panel) significance tests to search for a constant period spacing in the case of G\,29$-$38. The tests are applied to the pulsation periods in Table~\ref{table:G29_42_ModeID}.
A period spacing of $\sim 41$\,s is evident.}
    \label{GD_tests}
\end{figure}

In Fig.~\ref{FTG29_S42_ModeID}, we show the pulsation spectrum of G\,29$-$38 in terms of the periods.
The vertical red lines (blue lines) indicate the location of $\ell=1$ ($\ell= 2$) 
periods that produce the patterns of constant dipole and quadrupole period spacings.
We searched for a  constant period spacing in the data of G\,29$-$38  using the  Kolmogorov-Smirnov \citep[K-S;][]{1988IAUS..123..329K}, the inverse variance \citep[I-V;][]{1994MNRAS.270..222O}, and the Fourier Transform
\citep[F-T;][]{1997MNRAS.286..303H} significance tests. In the K-S
test, the quantity $Q$ is defined as the probability that the observed
periods are randomly distributed. Thus, any uniform or at least
systematically non-random period spacing in the period spectrum of the
star will appear as a minimum in $Q$. In the I-V test, a maximum of
the inverse variance will indicate a constant period spacing. Finally,
in the F-T test, we calculate the FT of a Dirac comb
function (created from a set of observed periods), and then we plot
the square of the amplitude of the resulting function in terms of the
inverse of the frequency. A maximum in the square of
the amplitude will indicate a constant period spacing.
Fig.~\ref{GD_tests} displays the results of applying the three 
significance tests to the period spectrum of G\,29$-$38. We adopted the full set of 57 periods of Table~\ref{table:G29_42_ModeID}.  The three tests point to the existence of a clear pattern of $\ell= 1$ constant period spacing of $\Delta \Pi\sim 41$\,s.

\begin{figure}
	\includegraphics[width=\columnwidth]{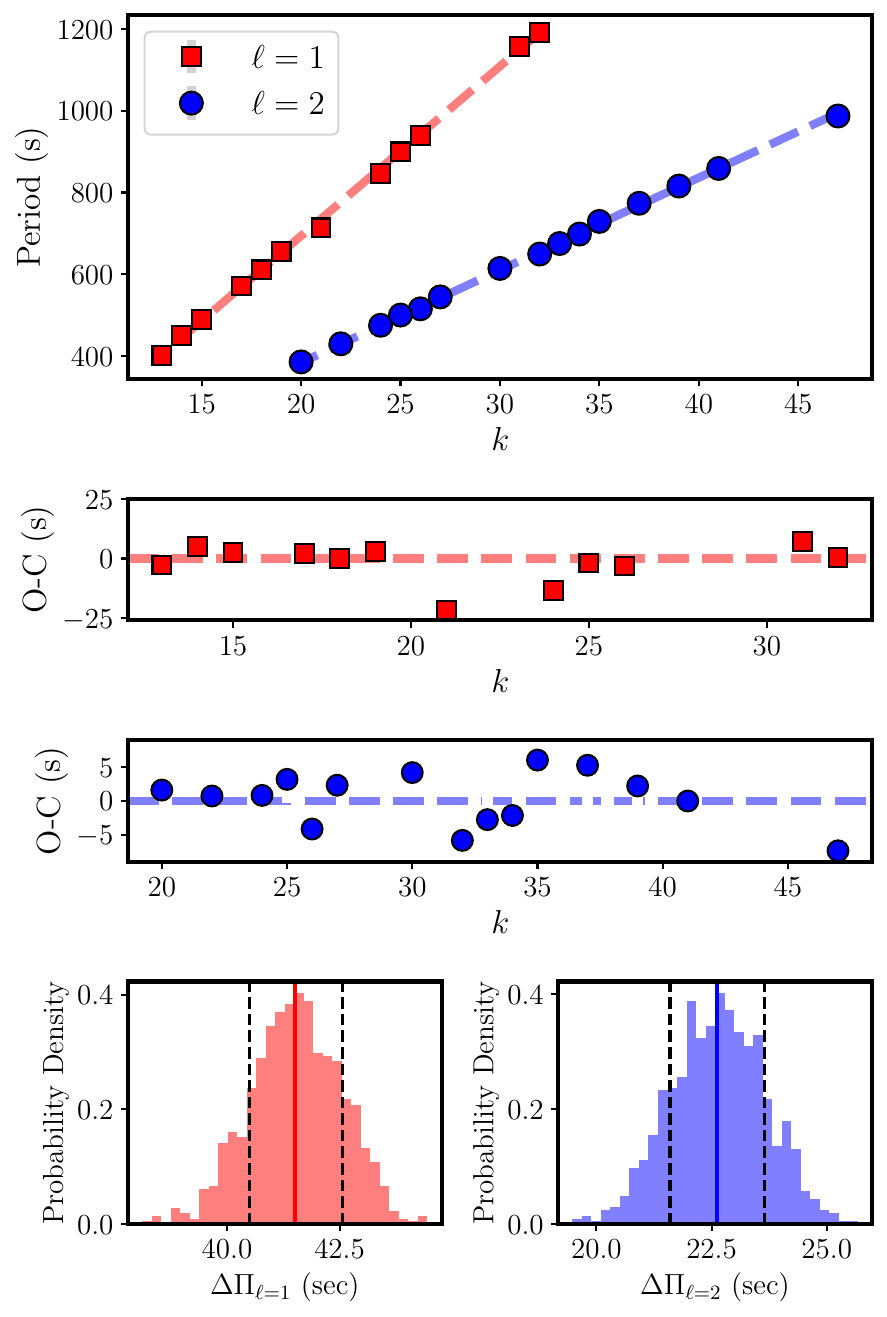}
    \caption{
    Linear least-squares fit the periods of G\,29$-$38 marked with filled red squares ($\ell=1$) and blue dots ($\ell=2$).
  The derived period spacing from this
  fit is $\Delta \Pi_{\ell=1}= 41.20$\,s and $\Delta \Pi_{\ell=2}= 22.58$\,s.  The residuals for $\ell=1$ modes (second panel) and $\ell=2$ modes (third panel)  of the period distribution relative to the mean period spacing.
  The fourth panel shows the distribution of the resulting fits for $\ell=1$ modes (left panel) and $\ell=2$ modes (right panel). The lower and upper bounds, which are shown with vertical dashed black lines are calculated by determining the 16th and 84th percentiles of each distribution. The derived mean period spacing corresponds to the 50th percentiles, which are shown with a vertical red line for $\ell=1$ modes and a vertical blue line for $\ell=2$ modes. 
  Note that the radial order $(k)$ assignation has been done arbitrarily, see Sect. \ref{APS} for more details on mode identification and the mean period spacing computations. 
     }
    \label{GD_dipole_dp}
\end{figure}

To derive a refined value of the period spacing, the identified 12 $\ell = 1$ and 15 $\ell = 2$ modes were used to obtain the mean period spacing through an LLS fit. 
We note that the uncertainties associated with the measurements might be underestimated because some of the pulsational modes are members of incomplete rotational triplets or quintuplets in which we cannot assess the central component ($m = 0$) of the modes.  
Therefore, to accurately assess the actual uncertainty, we performed fits on 1000 permutations of the periods as described in \cite{2019A&A...632A..42B} and \cite{2021A&A...651A.121U}. In each fit, we randomly assigned a value of $m \in \{-1,0,1\}$ for triplets and $m \in \{-2,-1,0,1,2\}$ for quintuplets to every observed mode and then adjusted to the intrinsic value of $m = 0$ using an assumed rotational splitting. The distribution of each fit is shown in the fourth panel of Fig.~\ref{GD_dipole_dp}. By calculating the standard deviation of the best-fit slopes, which amounts to 0.96\,s for dipole modes and 1.02\,s for quadrupole modes, we accounted for additional uncertainty. 
We obtain a period spacing  of $\Delta \Pi_{\ell =1 }= 41.20^{+1.98}_{-1.92}$\,s and  $\Delta \Pi_{\ell =2 }= 22.58^{+2.00}_{-2.05}$\,s.  
Our findings align with the value derived from the three significance tests conducted directly on the list of periods.
In the second and third panels of Fig.~\ref{GD_dipole_dp} we show the residuals
($\delta \Pi$) between the observed dipole periods ($\Pi_i^{\rm O}$) and the
periods derived from the mean period spacing ($\Pi_{\rm fit}$).   
The presence of two minima between  $k=20$ and 25 for $\ell =1$, and $k=25$ and 35 for $\ell =2$ in the distribution of residuals suggests the occurrence of mode-trapping effects inflicted by the presence of internal chemical transition regions. 


\begin{table*}
\setlength{\tabcolsep}{8.pt}
\renewcommand{\arraystretch}{1.05}
\centering
\caption{Detected frequencies, periods, and 
amplitudes (and their uncertainties) and the 
signal-to-noise ratio with identified pulsational modes along with rotational splittings.
}
\begin{tabular}{ccccccccr}
\hline
\noalign{\smallskip}
  $\nu$     &  $\Pi$  &  $A$   &  S/N  & $k$  & $\ell$  & $m$ & $\delta\nu$  \\ 
 ($\mu$Hz)  &  (s)    & (ppt)  &       &      &         &   & $\mu$Hz   \\
\noalign{\smallskip}
\hline
\noalign{\smallskip}

740.059* 	(21)	&  1351.244	(13)	&   0.742	(69)	&    6.9      &     &     &               &      \\
838.89*	    (18)	&  1192.051	(12)	&   0.857	(69)	&    7.9      &  32   &  1   &            &      \\
864.037* 	(19)	&  1157.358	(12)	&   0.801 	(69)	&    7.4      &  31   &  1   &            &      \\
1006.627*	(19)	&  993.417 	(11)	&   0.834	(69)	&    7.7      &  47   &  2   & +1         &  6.33    \\
1012.964*	(07)	&  987.202 	(10)	&   2.268	(69)	&    21.0     &  47   &  2   & 0          &      \\
1064.336*	(13)	&  939.553 	(11)	&   1.216	(69)	&    11.3     &  26   & 1    &            &      \\
1106.833   (17) 	&  903.478 	(14)	&   1.458 	(10)    &    11.7     &  25   & 1    & +1         &  5.11    \\
1111.944*	(05)	&  899.326 	(10)	&   11.84	(47)	&    109.8    &  25   & 1    &  0         &      \\
1115.196   (33)     &  896.712(27)      &   0.900   (13)    &     7.5     &  25   &  1   &  $-1$      &  3.25    \\
1151.511*	(06)	&  868.424 	(10)	&   2.801	(69)	&    25.9     &  41   &  2   &  $-2$      &  12.84    \\
1164.353*	(04)	&  858.846 	(10)	&   4.056	(69)	&    37.6     &  41   &  2   &  0         &      \\
1181.656*	(10)    &  846.270 	(10)	&   1.49	(69)	&    13.8     &  24   &  1   &            &      \\
1210.47* 	(02)	&  826.125 	(10)	&   8.668	(69)	&    80.4     &  39   &  2   & +2         &  15.28    \\
1225.755*	(30)    &  815.824 	(12)	&   1.085	(10)	&    10.1     & 39    &  2 &   0          &      \\
1232.854   (11)	    &  811.125 	(77)	&  2.197 	(10)    &    17.6     &  39   & 2  &  $-1$        &  7.10    \\
1279.511*	(17)	&  781.549 	(11)	&   0.914	(69)	&    8.5      &  37   &  2   & +2         &  13.10    \\
1292.603   (41)     &  773.646 (79)     &  4.240    (13)    &    35.3     &  37  &  2 &   0           &      \\
1298.883   (02)	    &  769.892 	(14)	&  10.544	(10)    &    84.4     &  37   &  2   &  $-1$      &  6.28    \\
1307.303   (15)     &  764.942 (66)     &  2.627    (13)    &    21.9      &  37   &  2  &  $-2$      &  8.42    \\
1371.426*	(12)	&  729.168 	(10)	&   1.315	(69)	&    12.2     &  35   &  2   &            &      \\
1401.587*	(18)	&  713.477 	(10)	&   0.85	(69)	&    7.9      &  21   &  1   &            &      \\
1431.995*	(25)	&  698.327 	(11)	&   0.612	(69)	&    5.7      &  34   &  2   &            &      \\
1475.167*	(10)    &  677.889 	(10)	&   1.627	(69)	&    15.1     &   33   & 2    & +1        &  6.17    \\
1481.34* 	(08)	&  675.065 	(10)	&   1.906	(69)	&    17.7     &   33   &  2   &  0        &      \\
1487.704*	(14)	&  672.177 	(10)	&   1.074	(69)	&    9.9      &   33   &   2  &  $-1$     &  6.36    \\
1522.859*	(02)	&  656.660 	(10)	&   9.462	(69)	&    87.7      &  19   &  1   &  +1       &  3.73    \\
1526.590   (05)	    &  655.054 	(22)	&  4.983	(10)    &  39.9       &  19   &  1   &   0        &      \\
1530.651*	(03)	&  653.317 	(10)	&   4.959	(69)	&    45.9     &  19   &  1   &  $-1$      &   3.91   \\
1539.918*	(11)	&  649.385 	(10)	&   1.467	(69)	&    13.6      &  32   &  2   &           &      \\
1628.166*	(01)	&  614.188 	(10)	&   16.122	(69)	&    149.4    &  30   &  2   &            &      \\
1633.792   (05)	    &  612.073 	(19)	&  5.018	(10)    &  40.2       &  18   &  1   &  +1        &  3.76    \\
1637.552   (05)	    &  610.667 	(19)	&  5.035	(10)    &  40.3       &  18   &  1   &  0         &      \\
1642.383   (09)	    &  608.871 	(36)	&  2.640	(10)    &  21.1       &  18   &  1   & $-1$       &  4.83    \\
1649.424*	(11)	&  606.272 	(10)	&   1.346	(69)	&    12.5     &       &     &             &      \\
1745.251   (18)   	&  572.983 	(62)	&  1.339	(10)    &  10.7       &  17   & 1    & +1         &  5.39    \\
1750.641   (36)	    &  571.219 	(11)    &  0.696	(10)    &  5.6        &  17   & 1    &  0         &      \\
1756.076   (09)	    &  569.451 	(29)    &  2.796	(10)    &  22.4       &  17   & 1    & $-1$       &  5.43    \\
1836.735   (11)	    &  544.444 	(34)    &  2.186	(10)    &  17.5       &  27   &  2   &            &      \\
1940.523*	(15)	&  515.325 	(10)	&   1.064	(69)	&    9.9      &  26   &  2   &            &      \\
1986.868   (11)     &  503.304 (26)     & 1.261     (13)    &  10.5       &  25   &  2   &  +2        &  5.44    \\
1992.310	(48)	&  501.930	(12)	&   0.687	(69)    &	6.3       &  25   &  2   &  +1        &  7.43    \\
1999.742*	(02)	&  500.065 	(10)	&   6.948	(69)	&    64.4      &  25   &  2   &  0        &      \\
2006.51* 	(09)	&  498.378 	(10)	&   1.811	(69)	&    16.8     &  25   &  2   &   $-1$     &   6.77   \\
2013.93* 	(12)	&  496.542 	(10)	&   1.274	(69)	&    11.8     &  25   &  2   &   $-2$     &   7.42   \\
2016.620   (26) 	&  495.879 (63)     &  2.476    (11)    &  23.1       &     &     &  -            &      \\
2045.91* 	(18)	&  488.780 	(10)	&   0.853	(69)	&    7.9       &  15   &  1   &           &      \\
2104.979*	(27)	&  475.064	(62)	&  0.809	(99)    & 	7.5       &  24   & 2    &            &      \\
2223.76* 	(04)	&  449.689 	(10)	&   3.499	(69)	&    32.4     &  14   & 1    &            &      \\
2327.068*	(18)	&  429.725 	(10)	&   0.881	(69)	&    8.2      &  22   & 2    &            &      \\
2492.399   (04)	    &  401.219 	(08)    &  5.216	(10)    &  41.7       &  13   & 1    & +1         &  4.78    \\
2497.176*   (17)	&  400.452 	(28)    &  1.455	(10)    &  11.7       &  13   & 1    &  0         &      \\
2502.278*	(07)	&  399.636 	(10)	&   2.345	(69)	&    21.7     &  13   & 1    & $-1$       &  4.80    \\
2594.995*	(21)	&  385.357 	(10)	&   0.741	(69)	&    6.9      &  20   &  2   &            &      \\
3754.433*	(17)	&  266.352 	(10)	&   0.893	(69)	&    8.3      &     &     &               &      \\

\noalign{\smallskip}
\hline
\noalign{\smallskip}
\multicolumn{7}{l}{The frequencies that are detected in sector 42 are unmarked.} \\
\multicolumn{7}{l}{The frequencies that are detected in sector 56 are marked with an asterisk.} \\
\end{tabular}
\label{table:G29_42_ModeID}
\end{table*}




\section{Evolutionary models}
\label{evolutionary_models}

The asteroseismological analysis presented in this work is based on 
full DA WD evolutionary models that consider the complete evolution of the progenitor stars. Specifically, the models adopted here are taken from \cite{2010ApJ...717..897A} generated with the {\tt LPCODE} evolutionary
code. {\tt LPCODE} computes the complete evolution of the WD progenitor 
from the main sequence, through the hydrogen and helium burning stages, the thermally pulsing and mass-loss stages on the AGB, and the WD cooling phase. Thus, these models are characterized by consistent chemical profiles for both the core and envelope. The models adopt the convection scheme ML2 with the mixing length parameter  $\alpha=1$ \citep{1971A&A....12...21B, 1990ApJS...72..335T}.
For details regarding the input physics and the evolutionary code, we refer the reader to the works of \cite{2010ApJ...717..897A} and \cite{2010ApJ...717..183R}.
These evolutionary tracks and models have been successfully employed in previous studies of hydrogen-rich pulsating WDs 
\citep[see. e.g.,][]{2010ApJ...717..897A,2017A&A...599A..21D,2018A&A...613A..46D, 2012MNRAS.420.1462R,2013ApJ...779...58R,2017ApJ...851...60R,2019MNRAS.490.1803R, 2022MNRAS.511.1574R}.
In this work, we consider a model grid of carbon-oxygen core WDs with stellar masses varying from 0.525 to 0.877$M_{\odot}$ with total helium content of $M_{\rm He}\sim 10^{-2}M_{\star}$ and hydrogen content ($M_{\rm H}$) varying from $10^{-4}$ to $10^{-9}M_{\star}$ (see Table~\ref{table:envelopes}). 
Once the models reach the ZZ Ceti instability strip,  non-radial $\ell=1, 2$ $g$-mode periods 
 are computed for each model. This is done employing the adiabatic version of the {\tt LP-PUL} pulsation code \citep{2006A&A...454..863C}.
 
\begin{figure}
	\includegraphics[width=\columnwidth]{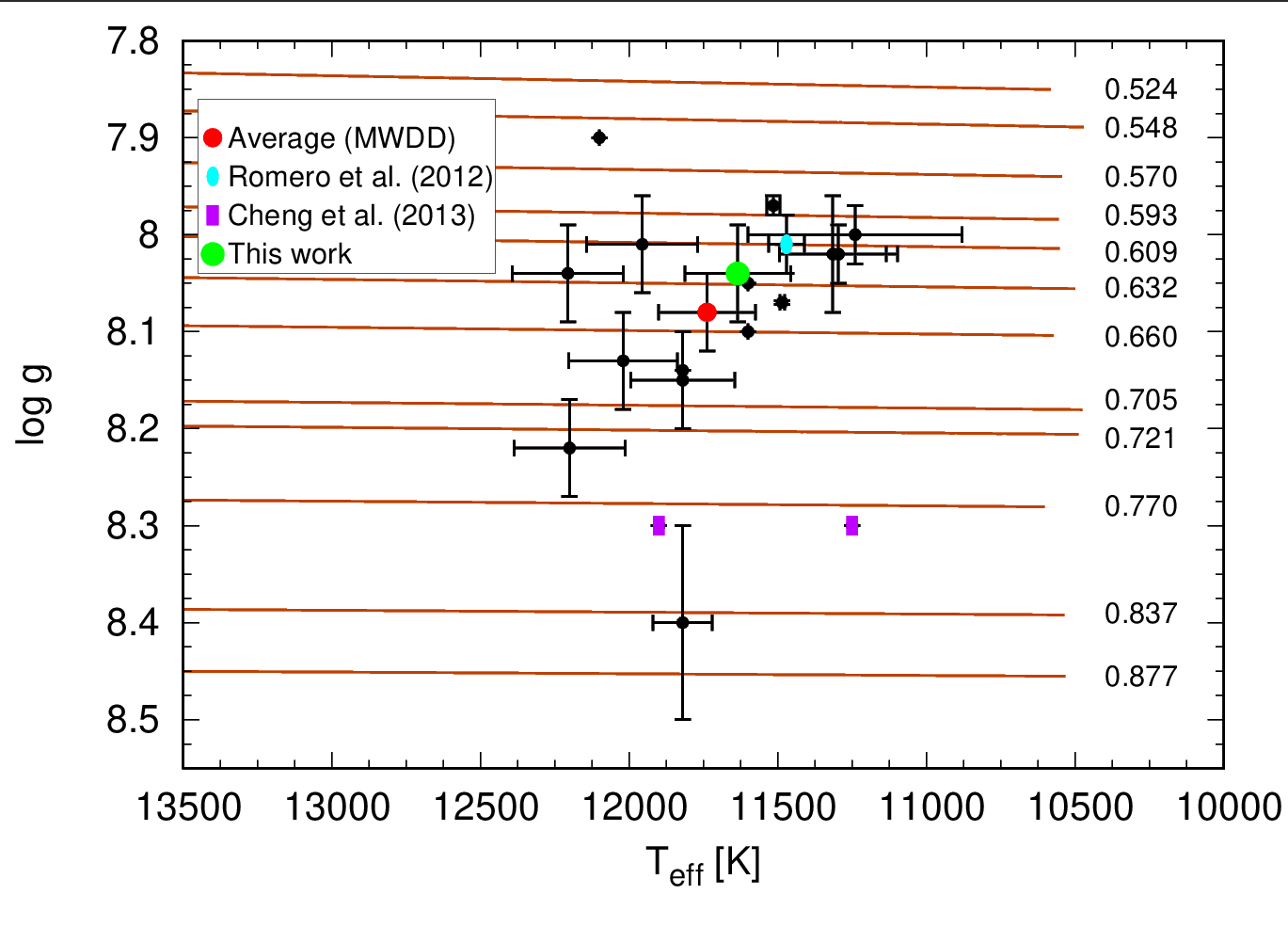}
    \caption{Determinations of $\log g $ and $T_{\rm eff}$ (black dots) for G\,29$-$38 from the Montreal White Dwarf Database (see the compilation in Table~\ref{basic-parameters-targets}) together with their average (red circle), our asteroseismic solution (green circle), and asteroseismic solutions from previous works (cyan ellipse and magenta rectangles). Superimposed on these,  we plot each of our canonical evolutionary sequences and their corresponding value of stellar mass in solar units (red lines)}
        \label{fig:teff-logg}
\end{figure}

From the previous spectroscopic determinations of $\log g$ and $T_{\rm eff}$, shown in Table~\ref{basic-parameters-targets}, we derived an average effective temperature and  $\log g$ of $11\,738\pm 162$\,K and $8.08\pm 0.04$, respectively.  In Fig.~\ref{fig:teff-logg} we show the spectroscopic measurements in the $T_{\rm eff}-\log g$ plane as well their average and previous asteroseismic determinations.  Superimposed on these, we also show our canonical evolutionary sequences\footnote{Sequences computed with the largest H content imposing that further evolution does not lead to hydrogen thermonuclear flashes on the WD cooling track.} and our best-fit model (see next section). By interpolating on our grid of evolutionary tracks,   we found that the average values of $T_{\rm eff}$ and $\log g$ of G\,29$-$38 are compatible with a WD model with $M_{\star}= 0.651\,M_{\odot}$ if the canonical H envelopes are assumed.  The total H content of our DA WD models is treated as a free parameter.

\section{Asteroseismic analysis}
\label{asteroseismological_modelling}

Our asteroseismological analysis consists in searching for the model that best matches the
pulsation periods of our target star, G\,29$-$38. To this end, we seek the theoretical model whose period spectrum minimizes a quality function defined as the average of
the absolute differences between theoretical and observed periods. This method has been successfully applied in previous works of the La Plata Stellar Evolution and Pulsation Research Group\footnote{\url{https://fcaglp.fcaglp.unlp.edu.ar/evolgroup/}} for a wide variety of classes of pulsators \citep[][and references therein]{2012MNRAS.420.1462R, 2013ApJ...779...58R,2017ApJ...851...60R,2019A&A...632A.119C,2022A&A...668A.161C,2022A&A...659A..30C}. 

 Before describing the seismological analysis,
we extract information on the stellar mass range of G\,29$-$38 using the observed period spacing
below.


\subsection{The stellar mass of G\,29$-$38 compatible with the observed period spacing}
\label{aps}


\begin{figure*}
	\includegraphics[width=clip,width=500pt]{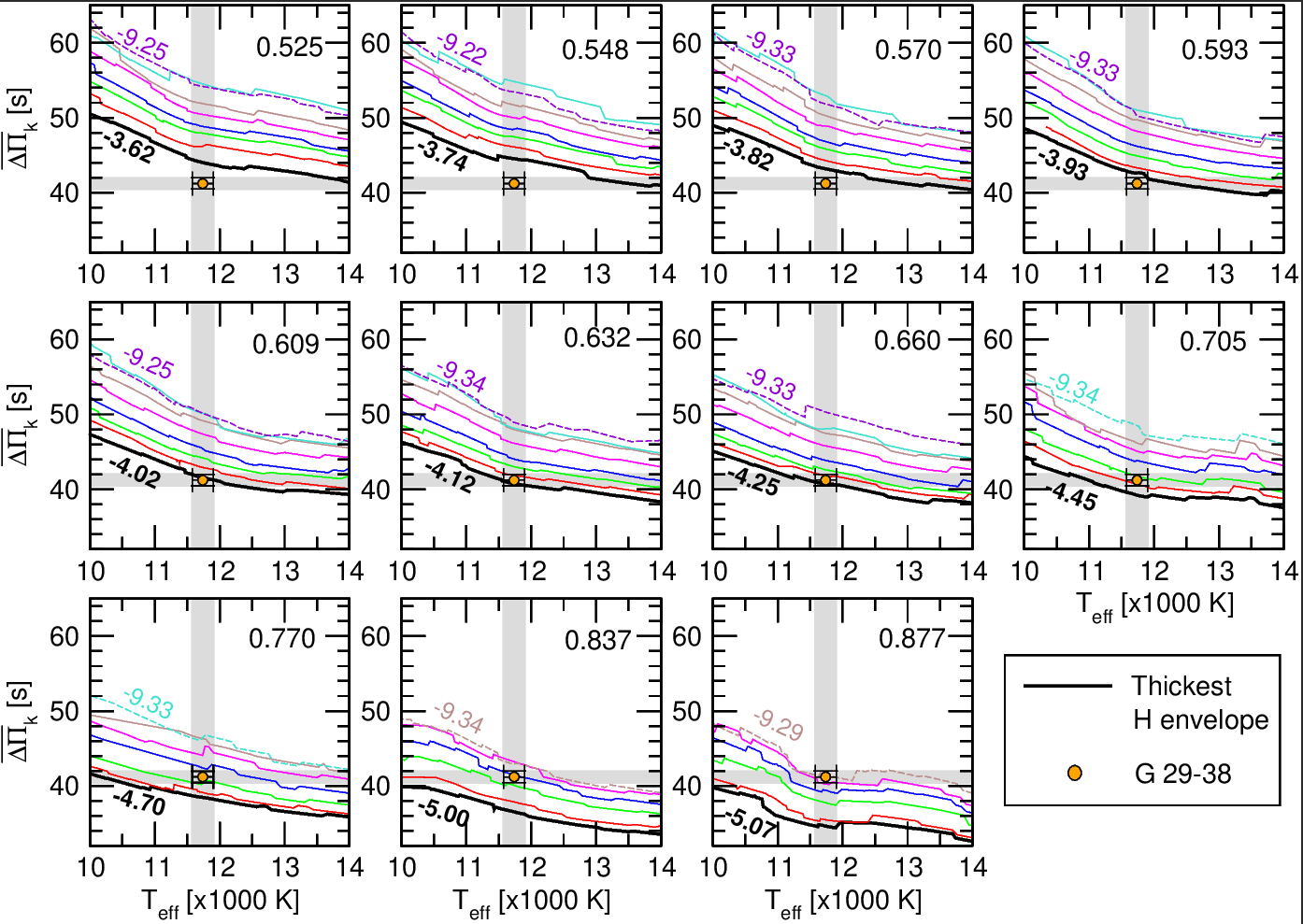}
    \caption{Average of the computed dipole ($\ell=1$) period spacings, $\overline{\Delta \Pi_k}$, in terms of the effective temperature, for different stellar masses in solar units (numbers at the top right corner of each panel) and thicknesses of the H envelope (see Table~\ref{table:envelopes}  for the specific values of $\log(M_{\rm H}/M_{\star})$) drawn with different colors.  In each panel, we include numbers along two curves, which correspond to the value of $\log(M_{\rm H}/M_{\star})$ for the thickest (black thick curves) and 
    the thinnest (violet, turquoise, and brown thin dashed curves, depending on $M_{\star}$) H envelopes for each stellar mass value. The location of G\,29$-$38 is emphasized with an orange circle with error bars ($T_{\rm eff}= 11\,738\pm162$\,K and $\Delta \Pi= 41.20^{+1.98}_{-1.92}$ s). The gray bands correspond to the uncertainties in $T_{\rm eff}$ and $\Delta \Pi$ of G\,29$-$38.}
\label{fig:acps}
\end{figure*}

\begin{table*}
\centering
\caption{The values of the stellar mass of our set of DA WD models 
(upper row) and the mass of H corresponding to 
the different envelope thicknesses considered for each stellar mass.
The second row shows the maximum value of the thickness of the H envelope
for each stellar mass (the ``canonical'' envelope thickness) according to our 
evolutionary computations. }
\begin{tabular}{r|cccccccccccc}
\hline
\hline
$M_{\star}/M_{\sun}$ &  0.525 &0.548 &0.570 & 0.593 &0.609 &0.632 & 0.660 & 0.705& 0.721& 0.770 & 0.837 & 0.877 \\
\hline
$\log(M_{\rm H}/M_{\star})$ & $-3.62$  & $-3.74$  & $-3.82$  & $-3.93$  & $-4.02$  & $-4.12$  & $-4.25$  & $-4.45$& $-4.50$ & $-4.70$  & $-5.00$  & $-5.07$ \\
\hline
                     & $-4.27$  & $-4.27$  & $-4.28$  & $-4.28$  & $-4.45$  & $-4.46$  & $-4.59$  & $-4.88$  & $\cdots$&$-4.91$  & $-5.41$  & $-5.40$ \\
                     & $-4.85$  & $-4.85$  & $-4.84$  & $-4.85$  & $-4.85$  & $-4.86$  & $-4.87$  & $-5.36$  &$-5.36$ &$-5.37$  & $-6.36$  & $-6.39$ \\
                     & $-5.35$  & $-5.35$  & $-5.34$  & $-5.34$  & $-5.35$  & $-5.35$  & $-5.35$  & $-6.35$  & $-6.43$&$-6.35$  & $-7.36$  & $-7.38$ \\
                     & $-6.33$  & $-6.35$  & $-6.33$  & $-6.33$  & $-6.34$  & $-6.34$  & $-6.35$  & $-7.35$  & $-7.34$&$-7.34$  & $-8.34$  & $-8.37$ \\
                     & $-7.34$  & $-7.33$  & $-7.34$  & $-7.34$  & $-7.33$  & $-7.35$  & $-7.33$  & $-8.34$  & $-8.33$&$-8.33$  & $-9.34$  & $-9.29$ \\
                     & $-8.33$  & $-8.33$  & $-8.31$  & $-8.33$  & $-8.33$  & $-8.33$  & $-8.33$  & $-9.34$  & $-9.24$&$-9.33$  &  $\cdots$  &   $\cdots$\\
                     & $-9.25$  & $-9.22$  & $-9.33$  & $-9.33$  & $-9.25$  & $-9.34$  & $-9.33$  &  $\cdots$   & $\cdots$&$\cdots$   &  $\cdots$   &   $\cdots$ \\
\hline 
\end{tabular}
\label{table:envelopes}
\end{table*}

A useful method to infer the stellar mass of pulsating WD stars is to compare
the measured period spacing ($\Delta \Pi$) with the average of the computed period spacings ($\overline{\Delta \Pi_{k}}$). This last quantity is calculated as $\overline{\Delta \Pi_{k}}= (n-1)^{-1} \sum_k \Delta \Pi_{k}$, where the ``forward'' period spacing ($\Delta \Pi_{k}$) is defined as $\Delta \Pi_{k}= \Pi_{k+1}-\Pi_{k}$ ($k$ being the radial order) and $n$ is the number of computed periods laying in the range
of the observed periods.  This method is more reliable for the estimation of the stellar mass than using the asymptotic period spacing, $\Delta \Pi_{\ell}^{\rm a}$ (Eq.~\ref{eq:1}), because, provided that the average of the computed period spacings is evaluated at the appropriate range of periods, the approach is valid for the regimes of short, intermediate, and long periods as well. When the average of the computed period spacings is taken over a range of periods characterized by high $k$ values, then the predictions of the present method become closer to those of the asymptotic period-spacing approach \citep[][]{2008A&A...478..175A}.  Note that these methods for assessing the stellar mass rely on the spectroscopic effective temperature, and the results are unavoidably affected by its associated uncertainty. The methods 
outlined above take full advantage of the fact that, generally,  the period spacing of pulsating WD stars primarily depends on the stellar mass and the effective temperature,  and very weakly on the thickness of the He envelope in the case of DBV stars, or the thickness of the O/C/He envelope in the case of GW~Vir stars \citep[see, e.g.,][]{1990ApJS...72..335T}. However, these methods cannot, in principle, be directly applied to DAV stars to infer the stellar mass, for which the period spacing depends simultaneously on $M_{\star}$, $T_{\rm eff}$, and $M_{\rm H}$ with comparable sensitivity, which implies the existence of multiple combinations of these three quantities that produce the same spacing of periods. For this reason, we will be able to provide only a possible \emph{range} of stellar masses for G\,29$-$38 on the basis of the period spacing.

We calculated the average of the computed period spacings for
$\ell= 1$, $\overline{\Delta \Pi_{k}}$,  in terms of the effective
temperature for all the stellar masses and H-envelope thicknesses considered (see Table~ \ref{table:envelopes}), and a period interval of $260-1400$\,s,
corresponding to the range of periods exhibited by G\,29$-$38. 
The results  are shown in Fig.~\ref{fig:acps}, where we depict  $\overline{\Delta
\Pi_{k}}$  for different stellar masses (specified at the right top corner of each panel) with curves of different colors according to the various values of $M_{\rm H}$.  For clarity, we have only labeled the thickest and the thinnest H envelope thickness value (for each stellar mass), with thick black and colored thin dashed curves, respectively. For the location of G\,29$-$38, indicated by a small orange circle with error bars, we considered the average spectroscopic effective temperature, $T_{\rm eff}= 11\,738\pm 162$\,K, and a period spacing $\Delta \Pi= 41.20^{+1.98}_{-1.92}$\,s. From an inspection of the plot, we conclude that according to the period spacing and $T_{\rm eff}$, the stellar mass of G\,29$-$38 should be between  $0.609\,M_{\odot}$ (with a thick H envelope of $\log(M_{\rm H}/M_{\star})= -4.02$) and $0.877\,M_{\odot}$ (with a very thin H envelope, of $\log(M_{\rm H}/M_{\star})= -9.29$). Although this constraint does not seem to be strong, it is actually precious because on the basis of $T_{\rm eff}$ and $\Delta \Pi$ (two measured quantities) we can rule out masses lower than $\sim 0.61\,M_{\odot}$ and possibly larger than $\sim 0.88\,M_{\odot}$ for G\,29$-$38. As we will see in the next section, most of the best solutions of the period fits are associated with WD models with masses in this range ($0.609 \lesssim M_{\star}/M_{\sun} \lesssim 0.877$). 

\subsection{Period-to-period fits}
\label{sect:period-fit}

\begin{table}
\caption{Period listing for each case described in section \ref{sect:period-fit}. 
For a clear understanding, we only show the integer part of each period.}
\begin{tabular}{lllll}
\hline
{\tt K98}   &{\tt T08}  & {\tt {\sl TESS}} &  {\tt K98+T08+TESS}   \\
\hline
 110&   &   &   110\\
    &  & 266& 266 \\
   &   & 385& 385\\
 400&    &   400 & 400\\
 &  431 & 429 &429\\
 &  & 449 & 449 \\
 &  & 475&  475\\
 &  &488&488\\
 &  &495&495\\
  500&   &  500 &500\\ 
  &   &515&515\\
    &   & 544& 544 \\
 552&   &   &552\\ 
   &   & 571 &571\\
   &  &606&606\\
   610&  & 610&  610\\ 
   
   & 614  &614 &614\\
 649&  &   649&649\\ 
 & 655 &655 &655\\
 &  &  675& 675 \\ 
 678&   681 &678 &678 \\
 &   &698&698\\
 &   &713&713\\
 730 &   &729   &729\\
   771&   776 & 773  &773 \\ 
 809&   815 &  815  &815\\
    &  835 & &835\\
    &    &846&846\\
 860&   &858   &858\\
894 &  &899& 899\\
 915    &920 & & 920 \\
   &  937 & 939 &939\\
 &     &987&987\\
 1147 & &&1147\\
 &  &1157&1157\\
 &   &1192&1192\\
 1240 &  &   &1240\\ 
 &  &1351&1351\\
  \hline
\end{tabular}
 \label{tab:cases}
\end{table}

We searched for the theoretical model that best fits each pulsation period of 
G\,29$-$38 individually.
In Table~\ref{tab:cases}, we summarized the periods list for the cases that were examined based on the findings from \citet{Kleinman1998} ({\tt K98}), \citet{thompson2008} ({\tt T08}) and {\sl TESS}.
We specifically examined the frequency spectrum of G\,29$-$38 and used the rotational triplets as input priors. We solely considered the central components (with $m=0$) of these triplets.
In cases where the rotational splitting did not provide a clear indication of the degree of modes, we assumed that those modes were either dipole or quadrupole modes.

To find the best seismic model, we evaluated the quality function:
\begin{equation}
\chi^2(M_{\star},M_{\rm H},T_{\rm eff})=\frac{1}{N} \sum_{i=1}^{N} {\rm min}[(\Pi_i^{\rm O}-\Pi_k^{\rm th})^2],   
\label{eq:quality}
\end{equation}
where $N$ is the number of detected modes, $\Pi_i^{\rm O}$ are the observed periods, and $\Pi_k^{\rm th}$ are the model periods.
The theoretical model that shows the minimum value in $\chi^2$ is adopted as our best-fit model.
We evaluated the quality function in our grid of models, that is 
for stellar masses in the range $0.525\leq M_{\star}/M_{\odot} \leq 0.877$, with effective temperature $10\, 000\,{\rm K} \leq T_{\rm eff}\leq 13\, 000$\,K, and varying the total hydrogen content $-9\lesssim \log(M_{\rm H}/M_{\star}) \lesssim -4$ , see Table~\ref{table:envelopes}.

\begin{figure}
		\includegraphics[width=\columnwidth]{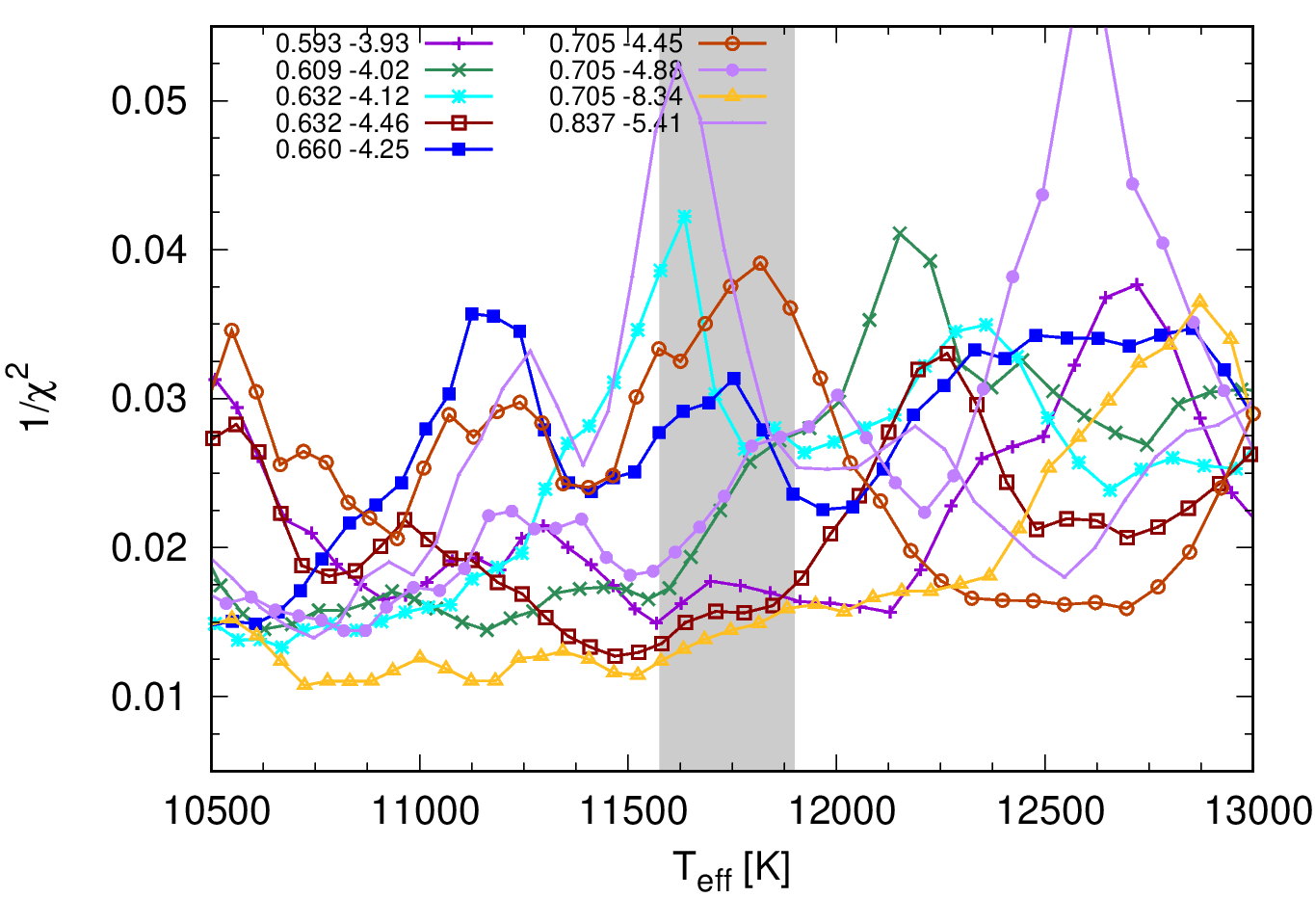}

    \caption{Inverse of the squared quality function $\chi$ in terms of the effective temperature for the best-fit models that agrees with {\sl Gaia} distance within 10\% (labeled here according to their respective $M_{\star}$, log$(M_{\rm H}/M_{\star})$ values). The grey region represents the averaged $T_{\rm eff}$ and its error from MWDD. For comparison, we also include the model that best matches the pulsation periods of G\,29$-$38.}
    \label{fig:quality}
\end{figure}

Our results are displayed in Table~\ref{tab:best-fits}. We found solutions that are compatible with recent
 spectroscopic determinations for $T_{\rm eff}$ and $\log g$,  which accounts for 3D corrections \citep{2013A&A...559A.104T}, as well as the astrometric distance provided by {\sl Gaia} (see next section). 
 Based on these results, it is most likely that G\,29$-$38 has a thick H-envelope. 
We are particularly interested in the case {\tt K98+T08+TESS} for which we found two potential solutions (seen as maxima in  Fig.~\ref{fig:quality}) with masses $0.632$ and $0.837\,M_{\odot}$ and similar effective temperature $\sim 11\,630$\,K. 
 Because of the disagreement with most of the spectroscopic $\log g$ determinations and the {\sl Gaia} distance, we regard the massive model as the less likely solution, despite it providing the best agreement between the theoretical and observed periods.
We prefer the solution characterized by   $M_{\star}/M_{\odot}=0.632$, $T_{\rm eff}=11\, 635$\,K,  and  log$(g)=8.048$ as our best-fit model. The location of this model in the $T_{\rm eff}~-~\log g$ diagram is displayed in Fig.~\ref{fig:teff-logg}, with a green circle. The stellar mass of the asteroseismological model found by means of the period-to-period fit analysis is in very good agreement with the results from our mean period spacing analysis but also with the most recent spectroscopic determinations. 
Combining the findings from {\tt K98}, {\tt T08} and {\sl TESS}, we fitted with 15 dipole modes with radial order $k$ in the range [7:30], and the remaining modes being quadrupoles with $k \in$ [2:48] with a value of the quality function of $\sigma=4.86$, or can be fitted with 19 dipole modes and $\sigma=5.74$. For the purpose of giving a quantitative evaluation of our best-fit model, we computed the 
average of the absolute period differences $\overline{\delta \Pi_i}=(\sum_{i=1}^{n}|\delta \Pi_i|)/n$, where $\delta \Pi_i=(\Pi_{\ell,k}-\Pi_i^{\rm o})$ and $n=38$. We found 
$\overline{\delta \Pi_i}=$ 3.97\,s, a value that is within our expectations given the large number of pulsation modes fitted (less than 1\,s per mode).

\begin{table*}
\caption{Potential best-fit models for each set of the periods considered for G\,29$-$38. Together with the basic stellar parameters, we list the number of pulsation periods associated with $\ell=1$ modes, the value of the quality function $\chi$, and the asteroseismic distance.}

\begin{tabular}{lllllllll}
\hline
 &$T_{\rm eff}$ (K)  &log $g$  & $M_{\rm WD}/M_{\odot}$ & $M_{\rm H}/M_{\rm WD}$ & $M_{\rm He}/M_{\rm WD}$ & $\ell=1$ & $\chi$ & $d$[pc]\\
\hline
K98               
                    & 11\,577  & 8.04  & 0.632 &7.58$\times 10^{-05}$	& 1.74$\times 10^{-02}$ &~~9  &3.74&17.45\\
     
T08                  

                 &11\,446  & 8.22  &0.721     &5.64$\times 10^{-10}$  &7.25$\times 10^{-03}$&~~4&3.24&15.30\\
{\sl TESS}          
                   &11\,635   & 8.04  &0.632  &7.58$\times 10^{-05}$  & 1.74$\times 10^{-02}$   &  15 &4.72 &17.54\\

{\tt TESS+K98+T08}  & 11\,620  & 8.39  &0.837  &3.91$\times 10^{-06}$	& 3.18$\times 10^{-03}$ &15 &4.36 &13.66\\
                    &11\,635   & 8.04  &0.632  &7.58$\times 10^{-05}$  & 1.74$\times 10^{-02}$   & 15  &4.86&17.54\\

\hline
\end{tabular}

\label{tab:best-fits}
\end{table*}

We give a global indicator of the quality of our asteroseismic fit that accounts for the free parameters and the value of the quality function, by  computing the Bayes Information Criterion \citep[BIC,][]{2000MNRAS.311..636K}:
\begin{equation}
{\rm BIC}= N_{\rm p} \left(\frac{\log N}{N} \right) + \log \sigma^2,
\end{equation}
where $N_{\rm p}$ is the number of free parameters of the models, $N$ is the number of observed periods to match and $\sigma$ the value of the quality function. The smaller the BIC value, the
better the quality of the fit. This criterion introduces a penalty term for an excess in the number of parameters
in the model. In our case, $N_{\rm p}= 3$ (stellar
mass, effective temperature, and thickness of the H envelope), 
$N=38$, and $\sigma^2=22.27$. 
We obtain ${\rm BIC}=1.47$, which means that our fit is good.

We assessed the internal uncertainties for the derived stellar mass, effective temperature, and surface gravity of the best-fit model by adopting the formula:
\begin{equation}
    \sigma_i^2=\frac{d_i^2}{(S-S_0)},
\end{equation}
where $\sigma_i$ refers to the uncertainty in each quantity, $S_0 \equiv \chi(M_*^0,M_H^0,T_{\rm eff}^0)$ is the minimum of $\chi$, the quality function, and $S$ is the value of $\chi$ when the parameter $i$ is changed by $d_i$ while the other parameters remain fixed \citep{1986ApJ...305..740Z, 2008MNRAS.385..430C, 2012MNRAS.420.1462R,2019A&A...632A.119C}. The parameter $d_i$ can be interpreted as the step in the
 grid of the quantity $i$. From the uncertainties in $M_{\star}$ and $T_{\rm eff}$, we derived the uncertainties in $\log g$, $L_{\star}$ and $R_{\star}$. We found $\sigma_{M_{\star}}=0.03~M_{\odot}$, $\sigma_{T_{\rm eff}}= 178$\,K, $\sigma_{\log g}=0.05$, $\sigma_{L_{\star}/L_{\odot}}=2\times 10^{-4}$ and $\sigma_{R_{\star}/R_{\odot}}=1\times 10^{-4}$. These errors are formal uncertainties inherent to the process of searching for the asteroseismological model.

\subsection{Asteroseismological distance}

We can estimate the asteroseismic distance for G\,29-38  based on the derived stellar parameters. From the effective temperature and gravity, we determined the absolute magnitude of our best-fit models in the {\sl Gaia} $G$ band (Koester priv. comm.). For the 0.632\,$M_{\odot}$ solution we find an absolute magnitude of $M_{\rm G}=11.839\pm 0.034$ mag. From the apparent magnitude obtained by {\sl Gaia} Data Release 3 (DR3) Archive\footnote{\url{https://gea.esac.esa.int/archive/}} for G\,29$-$38 ($m_{\rm G} =13.06$ mag),  we obtain an asteroseismic distance of $d = 17.54\pm 0.27$\,pc, and parallax of $\pi =56.9 \pm 1.$\,mas. 
An important aspect of validating our asteroseismic best-fit model is by comparing the asteroseismic distance with that obtained directly by {\sl Gaia}. We found an excellent agreement with the {\sl Gaia} distance \citep{Bailer-Jones2021}, which reports  of $d = 17.51^{+0.008}_{-0.007}$\,pc ($\pi = 57.097_{-0.023}^{+0.025}$\,mas). 
We repeated this process to each of the potential solutions.

\subsection{Comparison with previous works}

G\,29$-$38 has been the subject of several detailed period-to-period fit analyses in the past decades, based on the pulsation periods found by \cite{Kleinman1998} and \cite{thompson2008} (see Table~\ref{tab:solutions} for a summary of the most important stellar parameters derived in each study). 

The first detailed asteroseismic analysis for G\,29$-$38 was done by \cite{2009MNRAS.396.1709C} based on the mean period values\footnote{Periods assumed: 218, 283, 363,  400,  496,  614,  655,  770,  809,  859,  894,  1150,  1185,  1239\,s.}  detected by \cite{Kleinman1998}. These authors employed numerical models computed with the White Dwarf Evolutionary Code \citep[WDEC, see][ and references therein]{1990PhDT.......203W} in which they considered $T_{\rm eff}$, $M_{\star}$, $M_{\rm H}$ and $M_{\rm He}$ as free parameters, but a fixed core composition of 50\% $^{12}$C and 50\% $^{16}$O. By assuming all the observed pulsation periods as $\ell=1$ modes, the authors found an asteroseismic model characterized by $T_{\rm eff}=11\,700$\,K, $M_{\star}= 0.665\,M_{\odot}$ with a thin H-envelope.  The second asteroseismic analysis was performed by \cite{2012MNRAS.420.1462R} who adopted the same periods as in \cite{2009MNRAS.396.1709C} but their analysis was done adopting fully evolutionary models. The authors found a seismological solution for this star with $0.593\,M_{\odot}$, 11\,471\,K, and a very thin H-envelope of $4.67\times 10^{-10}M_{\star}$, with most observed pulsation periods fitted as  $\ell=2$ modes, except the 614\,s.


Finally, \cite{2013RAA....13.1438C} performed asteroseismological fits by adopting the period spectrum derived from \cite{thompson2008} and models from WDEC. These models resemble those from \cite{2009MNRAS.396.1709C} but with a different (fixed) core composition. The authors derived two best-fit solutions fitted with a mix of $\ell=1,2$ modes and characterized by $T_{\rm eff}=11\,900$\,K, $M_{\star}=0.790\,M_{\odot}$, $M_{\rm H}=10^{-4}M_{\star}$ and $T_{\rm eff}=11\,250$\,K, $M_{\star}=0.780\,M_{\odot}$, $M_{\rm H}=3.16\times 10^{-6}M_{\star}$. Both solutions have nearly the same mass and log$g$, but they differ in $T_{\rm eff}$ and the hydrogen content.

We found good agreement with previous asteroseismic determinations for $T_{\rm eff}$,  with maximum deviations of $\sim$ 3\%. 
In particular, our derived $M_{\star}$ show better agreement with that from \cite{2009MNRAS.396.1709C} and \cite{2012MNRAS.420.1462R}, with differences less than 7\% and larger differences when comparing with the results from \cite{2013RAA....13.1438C} --up to 25\%--. 
The comparison of other quantities such as the central abundance of C and O ($X_{\rm C}$, $X_{\rm O}$) or the thickness of the hydrogen envelope ($\log (1-M_{\rm H}/M_{\star})$) is more complex because of the different structures of the DA WD models and the different set of pulsation periods involved in each study.

\begin{table*}
\caption{Stellar parameters from previous asteroseismological studies for G\,29$-$38.  {\tt K98*} refers to the mean period values based on {\tt K98}. $X_{\rm C}$ and $X_{\rm O}$ refer to the central abundance of carbon and oxygen, respectively.}
\begin{tabular}{llllllllllll}
\hline
 & Dataset  &$T_{\rm eff}$ (K)& $M_{\star}$ ($M_{\odot}$) & $\log g$  & $M_{\rm H}/M_{\star}$ &  $M_{\rm He}/M_{\star}$& $\log(L/L_{\odot})$&$\log(R/R_{\odot})$&$X_{\rm C}$&$X_{\rm O}$  \\
 \hline

{\cite{2009MNRAS.396.1709C}}& {\tt K98*} & 11\,700 &0.665  & ... &1.0$\times 10^{-8}$  &1.$\times 10^{-2}$&...&...&0.500&0.500 \\
{\cite{2012MNRAS.420.1462R}} & {\tt K98*} & 11\,471 &0.593  &8.01 &4.6$\times 10^{-10}$  &2.39$\times 10^{-2}$ & -2.612&-1.901&0.283&0.704\\
 {\cite{2013RAA....13.1438C}}& {\tt T08} & 11\,900 &0.790  &8.30  &1.0$\times 10^{-4}$  &1.0$\times 10^{-2}$&...&...&0.200&0.800  \\
                            &           &11\,250  &0.780  &8.30  &3.1$\times 10^{-6}$  & 3.1$\times 10^{-3}$&...&...&0.200&0.800 \\
This work& {\tt K98+T98+TESS} &11\,635 &0.632&8.04&7.58$\times 10^{-5}$& $1.74\times 10^{-2}$&-2.594&-1.905&0.232& 0.755 \\ 
 \hline
\end{tabular}

\label{tab:solutions}
\end{table*}

\subsection{Uncertainties from the progenitor evolution}

Two primary approaches exist for conducting asteroseismic analysis of pulsating WD stars. One process involves constructing static stellar structures using parameterized luminosity and chemical profiles mildly based on stellar evolution outcomes \citep{2011ApJ...742L..16B,2014ApJ...794...39B,2019ApJ...871...13B,2014IAUS..301..285G,2016ApJS..223...10G,2017A&A...598A.109G}. While this method enables highly accurate fits, it may not fully align with current understanding of stellar evolution \citep[][]{2018ApJ...867L..30T,2019A&A...630A.100D} or with {\sl Gaia} astrometry \citep{2022RNAAS...6..244B}. The other approach, which is employed in this study, entails utilizing fully evolutionary models computed from the zero-age main sequence (ZAMS) to the ZZ Ceti stage \citep{2010ApJ...717..897A, 2012MNRAS.420.1462R}. It is worth noting, however, that these models are subject to uncertainties in the modeling of physical processes inside the stars. Past research has demonstrated that asteroseismic analysis of ZZ Ceti stars using fully evolutionary models can lead to deviations of up to 8\% in inferred values of $T_{\rm eff}$ and 5\% in $M_{\star}$, as well as up to two orders of magnitude in the mass of the H envelope \citep{2017A&A...599A..21D, 2018A&A...613A..46D}. These findings are primarily applicable to low-mass WDs, where uncertainties during prior evolution have a larger influence on the period spectrum of ZZ Ceti stars than in massive WDs \citep[$\gtrsim 0.8M_{\odot}$, see][]{2017A&A...599A..21D}. 

\section{Summary and Conclusions}
\label{conclusions}

This work presents a detailed astroseismological study of G\,29$-$38 based on short and ultra-short cadences {\sl TESS} observations. 
G\,29$-$38 was observed by {\sl TESS} in two sectors, sector 42 and sector 56, totaling 51 days. 
Using the high-precision photometry data, we identified 28 significant frequencies from sector 42 and 38 significant frequencies from sector 56. The oscillation frequencies have periods from $\sim$260\,s to $\sim$1400\,s and are associated with $g$-mode pulsations.
Additionally, we identified 30 combination frequencies per sector. 
Using the rotational frequency multiplets, we found four complete triplets and a quintuplet with a mean separation $\delta \nu_{\ell =1}$ = 4.67\,$\mu$Hz and $\delta \nu_{\ell =2}$ = 6.67\,$\mu$Hz, respectively, implying a rotation period of about 1.35 ($\pm 0.1$) days. 
This result is in line with what has been found by \citet{2017ApJS..232...23H}, 
who demonstrated that $0.51 - 0.73 M_{\odot}$ white dwarfs evolved from $1.7 - 3.0 M_{\odot}$ ZAMS progenitors, and have a mean rotation period of 1.46 d.

Based on the $\ell = 1$ and $\ell = 2$ modes defined by rotational triplets and quintuplets in conjunction with  statistical tests, 
we searched for a constant period spacing for $\ell = 1$ and $\ell = 2$ modes.  
Using solely {\sl TESS} observations, we identified 12 $\ell=1$ modes with radial order $k$ values ranging from 13 to 32 and 15 $\ell=2$ modes with  $k$ values between 20 and 47 as presented in Table \ref{table:G29_42_ModeID}.
We determined a constant period spacing of 41.20\,s for $\ell= 1$ modes and 22.58\,s for $\ell= 2$ modes, which are in good agreement with those inferred from the Kolmogorov-Smirnov, the inverse variance, and the Fourier transform statistical tests. We compared the constant period spacing obtained for the $\ell=1$ modes (41.20\,s) with that from our numerical models. Due to the intrinsic degeneracy of the dependence of $\Delta \Pi$ with $M_{\star}$, $T_{\rm eff}$ and $M_{\rm H}$ we were able to derive only a range for the stellar mass for G\,29$-$38 which is between $0.609\, M_{\odot}$ (with thick H envelope) and $0.877\, M_{\odot}$ (with thin H envelope). This analysis discards the existence of low-mass  ($< 0.609 M_{\odot}$)  solutions.

We combined the set of pulsation periods observed by  {\sl TESS}  and those from previous works \citep{Kleinman1998,thompson2008} and derived a complete set of pulsation periods for G\,29$-$38.
We applied exhaustive asteroseismic period-to-period analysis and derived an asteroseismological model with stellar mass $M_{\star}/M_{\odot}=0.632 \pm 0.03$, which is in good agreement with the value inferred from the period spacing analysis and also with the most recent spectroscopic determinations.
 Our results are in very good agreement with the asteroseismic results from \cite{2009MNRAS.396.1709C} and \cite{2012MNRAS.420.1462R}, regarding the derived $T_{\rm eff}$ and $M_{\star}$. 
 Finally, from the derived $T_{\rm eff}$ and $\log g$, we estimated the seismological distance 
of our best-fit model (17.54\,pc) that is in excellent agreement with that
 provided directly by {\sl Gaia} (17.51\,pc).



 \section*{Acknowledgements}

F.C.D.G.  acknowledges the financial support provided by FONDECYT grant No. 3200628. Additional support for F.C.D.G. and M.C. is provided by ANID's Millennium Science Initiative through grant ICN12\textunderscore009, awarded to the Millennium Institute of Astrophysics (MAS), and by ANID's Basal project FB10003.
Part of this work was supported by AGENCIA through the Programa de Modernización Tecnológica BID 1728/OC-AR, and by the PIP 112-200801-00940 grant from CONICET. M.C. is supported by FONDECYT Regular project No. 1231637. OT was supported by a FONDECYT project 321038.
MK acknowledges support from the National Science Foundation under grants AST-1906379 and AST-2205736, and NASA under grant 80NSSC22K0479. 
This research received funding from the European Research Council under the European Union’s Horizon 2020 research and innovation programme number 101002408 (MOS100PC) and 101020057 (WDPLANETS).

We gratefully acknowledge Prof. Detlev Koester for providing us with a tabulation of the absolute magnitude of DA WD models in the {\it Gaia} photometry.

This paper includes data collected with the {\sl TESS} mission, obtained from the MAST data archive at the Space Telescope Science Institute (STScI). Funding for the {\sl TESS} mission is provided by the NASA Explorer Program. 

This work has made use of data from the European Space Agency (ESA) mission {\sl Gaia} (\url{https://www.cosmos.esa.int/gaia}), processed by the {\sl Gaia} Data Processing and Analysis Consortium (DPAC, \url{https://www.cosmos.esa.int/web/gaia/dpac/consortium}). Funding for the DPAC has been provided by national institutions, in particular, the institutions participating in the {\sl Gaia} Multilateral Agreement. 

This research has made use of NASA's Astrophysics Data System Bibliographic Services, and the SIMBAD database, operated at CDS, Strasbourg, France.

\section*{Data Availability}
Data from {\sl TESS} is available at the MAST archive: \url{https://mast.stsci.edu/search/hst/ui/$#$/}. 

\bibliographystyle{mnras}
\bibliography{myref} 

\appendix

\section{Full frequency list}
\label{appendix}

\begin{table}
\setlength{\tabcolsep}{1.5pt}
\renewcommand{\arraystretch}{0.92}
\centering
\caption{Detected frequencies, periods, and 
amplitudes (and their uncertainties) and the 
signal-to-noise ratio from the  data of sector 42.}
\begin{tabular}{ccccr}
\hline
\noalign{\smallskip}
 $\nu$    &  $\Pi$  &  $A$   &  S/N   \\ 
  ($\mu$Hz)      &  (s)   & (ppt)   &   \\
\noalign{\smallskip}
\hline
\noalign{\smallskip}

1106.833 (17)	&  903.478 	(14)	&  1.458 	(10)  &  11.7  \\
1115.196 (33)   &  896.712  (27)   & 0.900 (13)  &    7.5        \\
1223.515 (27)	&  817.316 	(18)	&  0.908 	(10)  &  7.3   \\
1231.161 (15)	&  812.241 	(10)	&  1.662 	(10)  &  13.3  \\
1232.854 (11)	&  811.125 	(77)	&  2.197 	(10)  &  17.6 \\
1292.603 (41)   &  773.646 (79)   &  4.240 (13) &  35.3   &     \\
1298.883 (02)	&  769.892 	(14)	&  10.544	(10)  &  84.4  \\
1307.303 (15)    &  764.942 (66)  & 2.627 (13)	  &   21.9      \\
1474.048 (07)	&  678.403 	(35)	&  3.307	(10)  &  26.5  \\
1526.590 (05)	&  655.054 	(22)	&  4.983	(10)  &  39.9  \\
1530.251 (05)	&  653.487 	(23)	&  4.736	(10)  &  38.0  \\
1633.792 (05)	&  612.073 	(19)	&  5.018	(10)  &  40.2  \\
1637.552 (05)	&  610.667 	(19)	&  5.035	(10)  &  40.3  \\
1642.383 (09)	&  608.871 	(36)	&  2.640	(10)  &  21.1  \\
1745.251 (18)	&  572.983 	(62)	&  1.339	(10)  &  10.7  \\
1750.641 (36)	&  571.219 	(11)    &  0.696	(10)  &  5.6   \\
1756.076 (09)	&  569.451 	(29)    &  2.796	(10)  &  22.4  \\
1836.735 (11)	&  544.444 	(34)    &  2.186	(10)  &  17.5  \\
1986.868 (11)  	&  503.304 (26)   & 1.261 (13)   &  10.5         \\
2001.060 (02)	&  499.735 	(06)    &  9.904	(10)  &  79.2  \\
2016.620 (26) 	&  495.879 (63)    &  2.476 (11) &  23.1         \\
2110.152 (38)	&  473.899 	(86)    &  0.655	(10)  &  5.2   \\
2492.399 (04)	&  401.219 	(08)    &  5.216	(10)  &  41.7  \\
2497.176 (17)	&  400.452 	(28)    &  1.455	(10)  &  11.7  \\
2501.974 (18)	&  399.684 	(29)    &  1.373	(10)  &  11.0  \\
2747.582 (29)	&  363.956 	(39)    &  0.845	(10)  &  6.8   \\
3522.773 (07)	&  283.867 	(06)    &  3.263	(10)  &  26.1  \\
3639.341 (30)	&  274.775 	(23)    &  0.822	(10)  &  6.6   \\

\hline
\noalign{\smallskip}
 
1526.590 - 1298.883  &   227.829 (39)	&  4389.244 (76)	&  0.634 	(10)  &  5.1   \\
1633.792 - 1298.883 &   334.773 (31)	&  2987.092 (27)	&  0.809 	(10)  &  6.5   \\
4134.738 - 3790.969  &   344.944 (37)	&  2899.014 (31)	&  0.670 	(10)  &  5.4   \\
2001.060 - 1633.792  &   367.142 (31)	&  2723.741 (23)	&  0.812 	(10)  &  6.5   \\
2492.399 - 2001.060  &   491.278 (20)	&  2035.504 (85)	&  1.219 	(10)  &  9.8   \\
2001.060 - 1298.883  &   702.168 (12)	&  1424.159 (24)	&  2.103 	(10)  &  16.8  \\
2492.399 - 1633.792   &  858.687 (35) &  1164.568 (47)	&  0.714 (10)  &  5.7          \\
1986.868 - 1106.833 &   879.813 (38)	&  1136.605 (50)	&  0.650 	(10)  &  5.2   \\
2492.399 - 1298.883  & 1193.387 (26)	 &  837.951  (18)	&  0.934 (10)  &  7.5          \\
3522.773 - 1298.883 &    2223.717 (39)	&  449.697 	(81)    &  0.631	(10)  &  5.1   \\
2*1298.883            &  2598.084 (21)	&  384.898 	(32)    &  1.178 (10)  &  9.4      \\
1232.854 + 1530.251 &    2762.973 (29)	&  361.929 	(39)    &  0.849	(10)  &  6.8   \\
1298.883 + 1474.048  &     2772.942 (26)	&  360.627 	(34)    &  0.965 (10)  &  7.7      \\
1298.883 + 1526.590  &    2825.702 (25)	&  353.894 	(32)    &  0.973 (10)  &  7.8      \\
1292.603 + 1637.553    &    2930.010 (26)	&  341.295 	(31)    &  0.957	(10)  &  7.7   \\
1298.887 + 1637.552  &    2936.468 (29)	&  340.545 	(35)    &  0.844	(10)  &  6.8   \\
1298.887 + 1642.383  &    2941.162 (35)	&  340.001 	(42)    &  0.702	(10)  &  5.6   \\
1474.048 + 1526.590  &    3000.584 (36)	&  333.268 	(40)    &  0.699 (10)  &  5.6      \\
1298.883 + 1756.076  &    3054.962 (33)	&  327.336 	(35)    &  0.760 (10)  &  6.1      \\
1106.833 + 2001.060  &    3107.923 (19)	&  321.758 	(21)    &  1.265	(10)  &  10.1  \\
1530.251 + 1637.552  &    3167.776 (27)	&  315.678 	(28)    &  0.908 (10)  &  7.3      \\
1307.303 + 1986.868  &    3294.300 (27)	&  303.5546 (89)  & 0.94 (11)  & 8.8           \\
1298.883 + 2001.060 &    3300.127 (19)	&  303.018 	(18)    &  1.304	(10)  &  10.4  \\
1474.048 + 2001.060  &    3475.100 (24)	&  287.761 	(20)    &  1.046	(10)  &  8.4   \\
1526.590 + 2001.060 &    3527.709 (29)	&  283.470 	(24)    &  0.862	(10)  &  6.9   \\
1530.251 + 2001.060 &    3531.294 (31)	&  283.182 	(25)    &  0.799	(10)  &  6.4   \\
1642.383 + 2001.060  &    3643.330 (36)	&  274.474 	(27)    &  0.698	(10)  &  5.6   \\
1298.883 + 2492.399 &    3790.969 (37)	&  263.784 	(26)    &  0.679	(10)  &  5.4   \\
2*2001.060  &    4002.150 (44)	&  249.865 	(28)    &  0.571	(10)  &  4.6           \\   
1642.383 + 2492.399  &    4134.738 (40)	&  241.853 	(23)    &  0.630	(10)  &  5.0   \\   
                                                                                        
\noalign{\smallskip}
\hline
\end{tabular}
\label{table:S42_Flist}
\end{table}

\begin{table}
\setlength{\tabcolsep}{1.5pt}
\renewcommand{\arraystretch}{0.92}
\centering
\caption{Detected frequencies, periods, and 
amplitudes (and their uncertainties) and the 
signal-to-noise ratio from the  data of sector 56.}
\begin{tabular}{ccccr}
\hline
\noalign{\smallskip}
  $\nu$    &  $\Pi$  &  $A$   &  S/N   \\ 
  ($\mu$Hz)      &  (s)   & (ppt)   &   \\
\noalign{\smallskip}
\hline
\noalign{\smallskip}

740.059 	(21)	&  1351.244	(13)	&   0.742	(69)	&    6.9         \\
838.89	    (18)	&  1192.051	(12)	&   0.857	(69)	&    7.9          \\
864.037 	(19)	&  1157.358	(12)	&   0.801 	(69)	&    7.4          \\
1006.627	(19)	&  993.417 	(11)	&   0.834	(69)	&    7.7          \\
1012.964	(07)	&  987.202 	(10)	&   2.268	(69)	&    21.0         \\
1064.336	(13)	&  939.553 	(11)	&   1.216	(69)	&    11.3         \\
1111.944	(05)	&  899.326 	(10)	&   11.84	(47)	&    109.8        \\
1112.643	(05)	&  898.761 	(10)	&   5.144	(74)	&    47.7         \\
1151.511	(06)	&  868.424 	(10)	&   2.801	(69)	&    26.0         \\
1164.353	(04)	&  858.846 	(10)	&   4.056	(69)	&    37.6         \\
1181.656	(10)    &  846.270 	(10)	&   1.49	(69)	&    13.8         \\
1210.47 	(02)	&  826.125 	(10)	&   8.668	(69)	&    80.4         \\
1225.755	(30)    &  815.824 	(12)	&   1.085	(10)	&    10.1         \\
1279.511	(17)	&  781.549 	(11)	&   0.914	(69)	&    8.5          \\
1371.426	(12)	&  729.168 	(10)	&   1.315	(69)	&    12.2         \\
1401.587	(18)	&  713.477 	(10)	&   0.85	(69)	&    7.9          \\
1431.995	(25)	&  698.327 	(11)	&   0.612	(69)	&    5.7          \\
1475.167	(10)    &  677.889 	(10)	&   1.627	(69)	&    15.1         \\
1481.34 	(08)	&  675.065 	(10)	&   1.906	(69)	&    17.7         \\
1487.704	(14)	&  672.177 	(10)	&   1.074	(69)	&    10.0          \\
1522.859	(02)	&  656.660 	(10)	&   9.462	(69)	&    87.7          \\
1530.651	(03)	&  653.317 	(10)	&   4.959	(69)	&    45.9         \\
1539.918	(11)	&  649.385 	(10)	&   1.467	(69)	&    13.6          \\
1628.166	(01)	&  614.188 	(10)	&   16.122	(69)	&    149.4        \\
1649.424	(11)	&  606.272 	(10)	&   1.346	(69)	&    12.5         \\
1940.523	(15)	&  515.325 	(10)	&   1.064	(69)	&    9.9         \\
1992.310	(48)	&  501.930	(12)	&   0.687	(69)    &	6.3          \\
1999.742	(02)	&  500.065 	(10)	&   6.948	(69)	&    64.4          \\
2006.51 	(09)	&  498.378 	(10)	&   1.811	(69)	&    16.8         \\
2013.93 	(12)	&  496.542 	(10)	&   1.274	(69)	&    11.8          \\ 
2045.91 	(18)	&  488.780 	(10)	&   0.853	(69)	&    7.9           \\
2104.979	(27)	&  475.064	(62)	&  0.809	(99)    & 	7.5           \\
2223.76 	(04)	&  449.689 	(10)	&   3.499	(69)	&    32.4         \\
2327.068	(18)	&  429.725 	(10)	&   0.881	(69)	&    8.2          \\
2492.19 	(07)	&  401.254 	(10)	&   2.37	(69)	&    22.0         \\
2497.184	(15)	&  400.451 	(10)	&   1.061	(69)	&    9.8          \\
2502.278	(07)	&  399.636 	(10)	&   2.345	(69)	&    21.7         \\
2594.995	(21)	&  385.357 	(10)	&   0.741	(69)	&    6.9          \\
3754.433	(17)	&  266.352 	(10)	&   0.893	(69)	&    8.3          \\


\hline
\noalign{\smallskip}

1628.166 - 1522.859 &  105.262 	(14)	&  9500.105	(1.26)	&   1.075	(69)	&    10.0         \\
1522.859 - 1210.47  &  312.364 	(11)	&  3201.393	(11)	&   1.361	(69)	&    12.6         \\
1475.167 - 1111.944 &  363.789 	(19)	&  2748.846	(14)	&   0.82	(69)	&    7.6           \\
1530.651 - 1164.353 & 366.346 	(18)	&  2729.660	(13)	&   0.886	(69)	&    8.2          \\
1999.742 - 1628.166 &  371.551 	(08)	&  2691.421	(15)	&   2.041	(69)	&    18.9         \\
1628.166 - 1210.47  &  417.708 	(05)	&  2394.017	(12)	&   3.247	(69)	&    30.1          \\
1999.742 - 1522.859 & 476.841 	(08)	&  2097.135	(13)	&   1.935	(69)	&    17.9         \\
1628.166 - 1111.944 &  516.212 	(18)	&  1937.189	(16)	&   0.866	(69)	&    8.0          \\
371.551 + 417.708 &   789.277 	(14)	&  1266.982	(12)	&   1.07	(69)	&    9.9             \\
417.708 + 476.841 &   894.484 	(21)	&  1117.963	(12)	&   0.729	(69)	&    6.8             \\
838.89 + 894.484  &   1733.431	(23)	&  576.891 	(10)	&   0.66	(69)	&    6.1             \\ 
1111.944 + 1064.336 & 2176.203	(20)    &  459.516 	(10)	&   0.776	(69)	&    7.2          \\
1111.944 + 1210.47 & 2322.685	(15)	&  430.536 	(10)	&   1.02    (69)	&    9.3             \\
1111.944 + 1522.859   & 2634.85 	(15)	&  379.528 	(10)	&   1.06	(69)	&    9.8         \\
1111.944 + 1530.651   & 2642.705	(11)	&  378.400 	(10)	&   1.42	(69)	&    13.2        \\
1111.944 + 1628.166   & 2740.103	(11)	&  364.950 	(10)	&   1.455	(69)	&    13.5         \\
1628.166 + 1210.47 & 2838.662	(10)    &  352.279 	(10)	&   1.573	(69)	&    14.6         \\
1522.859*2 &   3045.74 	(23)	&  328.327 	(10)	&   0.674	(69)	&    6.2                     \\  
1111.944 + 1999.742 & 3111.529	(17)	&  321.385 	(10)	&   0.918	(69)	&    8.5         \\
312.364 + 2838.662 &   3151.02 	(08)	&  317.358 	(10)	&   1.956*	(69)	&    18.1            \\
1530.651 + 1628.166 &   3158.84 	(21)	&  316.572 	(10)	&   0.737	(69)	&    6.8         \\
1628.166*2 &   3256.321	(12)	&  307.095 	(10)	&   1.283	(69)	&    11.9                    \\
1111.944 + 2223.76 & 3335.497	(11)	&  299.805 	(10)	&   1.36	(69)	&    12.6          \\
1522.859 + 1999.742 &   3522.785	(08)	&  283.866 	(10)	&   1.921	(69)	&    17.8        \\   
1628.166 + 1999.742 &  3627.914	(13)	&  275.641 	(10)	&   1.214	(69)	&    11.3         \\
1628.166 + 2223.76 & 3851.865	(18)	&  259.615 	(10)	&   0.875	(69)	&    8.1         \\
1628.166 + 2502.278 & 4130.428	(21)	&  242.106 	(10)	&   0.722	(69)	&    6.7           \\
2223.76*2  &   4447.332	(20)    &  224.854 	(10)	&   0.76	(69)	&    7.0                     \\


\hline
\end{tabular}
\label{table:S56_Flist}
\end{table}

\bsp	
\label{lastpage}
\end{document}